\documentclass[prc,notitlepage]{revtex4-1}
\usepackage{color,graphicx}
\usepackage{soul}

\begin{document}

\title{Dilepton emission in high-energy heavy-ion collisions with viscous hydrodynamics}

\author{Gojko Vujanovic$^1$, Clint Young$^2$, Bj\"orn Schenke$^3$, Ralf Rapp$^4$, Sangyong Jeon$^1$, and Charles Gale$^1$}
\affiliation{$^1$Department of Physics, McGill University, 3600 University Street, Montreal, QC H3A 2T8, Canada}
\affiliation{$^2$School of Physics \& Astronomy, University of Minnesota, Minneapolis, MN 55455, USA}
\affiliation{$^3$Physics Department, Bldg. 510A, Brookhaven National Laboratory, Upton, NY 11973, USA}
\affiliation{$^4$Cyclotron Institute and Department of Physics \& Astronomy, Texas A\&M University, College Station, TX 77843-3366, USA}

\date{\today}

\begin{abstract}
The invariant mass spectrum and the elliptic flow of lepton pairs produced in relativistic heavy-ion collisions at RHIC are studied with viscous hydrodynamics. The effects of viscous corrections on dilepton observables are explored. The lepton pairs originating from charm quarks evolving in the viscous background are seen to be  a good probe of quark energy loss and gain, as quantified by the dilepton spectrum and by the dilepton elliptic flow.
\end{abstract}

%\keywords{}

\maketitle

%%%%%%%%%%%%%%%%%%%%%%%%%%%%%%%%%%%%%%%%%%%%%%%%%%%%%%%%%%%%%%%%%%%%%%%%%%%%%%%%%%%%%%%%%%%%%%%%%%%%%%%%%%%%%%%%%%%%%%%%%%%%%%%%%%%%%%%%%%%%%%%%
\section{Introduction}%%%%%%%%%%%%%%%%%%%%%%%%%%%%%%%%%%%%%%%%%%%%%%%%%%%%%%%%%%%%%%%%%%%%%%%%%%%%%%%%%%%%%%%%%%%%%%%%%%%%%%%%%%%%%%%%%%%%%%%%%%
\label{intro}%%%%%%%%%%%%%%%%%%%%%%%%%%%%%%%%%%%%%%%%%%%%%%%%%%%%%%%%%%%%%%%%%%%%%%%%%%%%%%%%%%%%%%%%%%%%%%%%%%%%%%%%%%%%%%%%%%%%%%%%%%%%%%%%%%%
%%%%%%%%%%%%%%%%%%%%%%%%%%%%%%%%%%%%%%%%%%%%%%%%%%%%%%%%%%%%%%%%%%%%%%%%%%%%%%%%%%%%%%%%%%%%%%%%%%%%%%%%%%%%%%%%%%%%%%%%%%%%%%%%%%%%%%%%%%%%%%%%

The collision of large nuclei at relativistic energies constitutes the only practical way to heat and compress nuclear matter in the laboratory. Therefore, a vibrant experimental program that aims to elucidate the bulk properties of hot and dense strongly-interacting matter is being pursed at  several accelerator facilities around the world, notably at the Relativistic Heavy Ion Collider (RHIC, at Brookhaven National Laboratory) and at the Large Hadron Collider (LHC, at CERN). This  enterprise has several ambitious goals, but two of them are to map out of the phase diagram of QCD (QuantumChromoDynamics: the theory of the strong interaction), and to determine the transport coefficients of QCD, which regulate departures from equilibrium. From the volume of data acquired over the last thirty years, together with the theoretical approaches used to interpret them, a ``standard picture'' is emerging: the nuclear initial states interact strongly leading to rapid apparent thermalization, followed by a period of quasi-ideal hydrodynamic evolution which lasts until kinetic freeze-out.  The details of this scenario may differ according to approaches - and such details  will have empirical consequences - but it is fair to write that the global picture currently appears robust. The endurance of this conceptual framework owes much to the modelling success of relativistic hydrodynamics \cite{Gale:2013da}. 

Much of the data measured at RHIC and at the LHC consist of hadronic particles which mostly reflect the final stages of the interacting system. Electromagnetic probes, on the other hand,  have the potential to provide an unambiguous measurement of the interior dynamics in heavy-ion collisions, owing to the fact that their interaction is dictated by $\alpha_{\rm EM}$ (the electromagnetic fine-structure constant), and that $\alpha_{\rm EM} \ll \alpha_{\rm S}$ (where $\alpha_{\rm S} = g^2/ 4 \pi$, with $g$  being the strong interaction coupling constant). The photons | real and virtual | produced in the hot and dense medium are thus penetrating probes essentially impervious to final-state interactions and as such can reveal details of the underlying particle distributions, including the degree of departure from thermal equilibrium. It is important however that precise measurements be accompanied by equally precise modelling. Indeed, the last few years have seen much progress in the simulation of heavy-ion collisions with relativistic hydrodynamics. More specifically, 3+1D relativistic viscous hydrodynamics models are available and have been used to characterize the matter formed in the relativistic collisions of large nuclei. The current standard input to those simulation tools are hydrodynamic equations derived up to second order in flow velocity gradients for conformal and non-conformal fluids \cite{Baier:2007ix,Betz:2009zz}. This degree of theoretical and modelling sophistication, together with progress in the quantitative description of the initial states, has brought closer one of the goals described earlier: the extraction of a shear viscosity coefficient for QCD \cite{Gale:2012rq}.  

Going back to electromagnetic radiation, the real photon spectrum has also shown sensitivity to a finite shear viscosity coefficient, as well as to the morphology of the initial states \cite{dion-paquet-schenke-young-jeon-gale,Chatterjee:2011dw,Shen:2013cca}. They are therefore observables capable of carrying information from the earliest moments of the collision. The drawback is that, as photons and leptons are emitted throughout the space-time history of the nuclear collision,  precise knowledge of the emission rates and the spatial and temporal evolution is required to interpret the measured signal. 
Dileptons offer the same penetrating advantages as real photons, but their production is suppressed by an extra  factor of $\alpha_{\rm EM}$. However, lepton pairs have an additional degree of freedom, as the pair's invariant mass and three-momentum are independent. 

The goal of this article is to explore the sensitivity of the virtual photon spectrum - through its conversion to lepton pairs - to a nonzero value of the shear viscosity to entropy density ratio, $\eta/s$. As is the case for photons, a good knowledge of the different sources of dileptons is required in order to extract meaningful quantitative information. We will concentrate on dileptons produced in nuclear collisions at full RHIC energy,  $\sqrt{s} = 200$~A GeV.  There, dileptons from high-temperature QCD processes, dileptons from the hot hadronic medium, and dileptons from charm decays are each expected to dominate in different ranges of invariant mass. We include these sources, and study the effect of viscosity on the final lepton pair spectrum, which includes the coefficient of elliptic flow, $v_2$. 

In Section \ref{rates}, we first discuss a formulation of relativistic hydrodynamics that incorporates a non-zero coefficient of shear viscosity, and we highlight the differences between the inviscid and viscous evolutions.  Then, the derivation of ideal rates and their viscous corrections for both the quark-gluon plasma phase and the hadronic phase are summarized. The contribution to the dilepton yields in the intermediate mass range from charm decays, and its modification in a heavy-ion collision, is discussed in Section \ref{charm}. The dilepton production in the low and intermediate mass regions is discussed in detail in Section \ref{production}, with emphasis on how viscous corrections change observables. After a complete hydrodynamic simulation of the Au-Au collisions, we compare our yields against recent experimental data from the STAR collaboration at RHIC. Finally, we conclude in Section \ref{conclusion}.

%%%%%%%%%%%%%%%%%%%%%%%%%%%%%%%%%%%%%%%%%%%%%%%%%%%%%%%%%%%%%%%%%%%%%%%%%%%%%%%%%%%%%%%%%%%%%%%%%%%%%%%%%%%%%%%%%%%%%%%%%%%%%%%%%%%%%%%%%%%%%%%%
\section{Ideal thermal rates and their viscous corrections}%%%%%%%%%%%%%%%%%%%%%%%%%%%%%%%%%%%%%%%%%%%%%%%%%%%%%%%%%%%%%%%%%%%%%%%%%%%%%%%%%%%%%
\label{rates}%%%%%%%%%%%%%%%%%%%%%%%%%%%%%%%%%%%%%%%%%%%%%%%%%%%%%%%%%%%%%%%%%%%%%%%%%%%%%%%%%%%%%%%%%%%%%%%%%%%%%%%%%%%%%%%%%%%%%%%%%%%%%%%%%%%
%%%%%%%%%%%%%%%%%%%%%%%%%%%%%%%%%%%%%%%%%%%%%%%%%%%%%%%%%%%%%%%%%%%%%%%%%%%%%%%%%%%%%%%%%%%%%%%%%%%%%%%%%%%%%%%%%%%%%%%%%%%%%%%%%%%%%%%%%%%%%%%%
\subsection{Viscous relativistic fluid dynamics}

Before deriving the viscous correction to the emission rates of electromagnetic radiation, it is appropriate to summarize the effect of shear viscosity on the bulk dynamics. In this paper, we shall consider no other transport coefficients. As reviews can be found in the recent literature \cite{Hirano:2008hy,Heinz:2009xj,Gale:2013da}, this summary may be brief. In the conformal Israel-Stewart formalism \cite{Israel:1979wp}, the stress-energy tensor is usually expressed as:
\begin{eqnarray}
T^{\mu \nu} = T_{\rm ideal}^{\mu \nu} + \pi^{\mu \nu}
\end{eqnarray}
where $T^{\mu \nu}_{\rm ideal} = \left( \epsilon + {\mathcal P}\right) u^\mu u^\nu - {\mathcal P} g^{\mu \nu}$ is the part of the stress-energy tensor that is unaffected by viscous corrections. The metric tensor is $g_{\mu \nu} = g^{\mu \nu} = {\rm diag} (1, -1, -1, -1)$ and the flow velocity is $u^\mu = ( \gamma, \gamma {\bf v})$, which reduces to $u^\mu = (1, 0, 0, 0)$ in the fluid rest frame. The quantities $\epsilon$ and ${\mathcal P}$ are the energy density and the pressure, respectively.  To second order in flow velocity gradients,  the  equations that dictate the hydrodynamic evolution are
\begin{eqnarray}
\partial_\mu T^{\mu \nu} &=& 0\ \nonumber \\
\Delta^\mu_\alpha \Delta^\nu_\beta u^\sigma \partial_\sigma \pi^{\alpha \beta} &=& - \frac{1}{\tau_\pi} \left( \pi^{\mu \nu} - S^{\mu \nu} \right) - \frac{4}{3} \pi^{\mu \nu} \left(\partial_\alpha u^\alpha\right)
\label{EOM}
\end{eqnarray}
The viscous part of the stress-energy tensor to first order in flow velocity gradients (the Navier-Stokes limit) is $S^{\mu \nu} = \eta \left(\nabla^\mu u^\nu + \nabla^\nu u^\mu - \frac{2}{3} \Delta^{\mu \nu} \nabla_\alpha u^\alpha \right)$ and  $\tau_\pi$ is the shear relaxation time. The local three-metric and 
spatial 
derivative are $\Delta^{\mu \nu} = g^{\mu \nu} - u^\mu u^\nu$ and $\nabla^\mu = \Delta^{\mu \nu} \partial_\nu$, respectively, and $\eta$ is the coefficient of shear 
viscosity.

The dynamics of relativistic heavy ion collisions make it especially advantageous to work in so-called hyperbolic coordinates, such that the coordinate transformation is $x^\mu = (t, x, y, z) \to (\tau, x, y, \eta_s)$, with $\tau = \sqrt{t^2 - z^2}$ and $\eta_s = (1/2)  \ln\left[(t+z)/(t-z)\right]$: the space-time rapidity. In addition, it is straightforward to show that $t = \tau \cosh \eta_s$, $z = \tau \sinh \eta_s$, and that $g_{\mu \nu} = {\rm diag}(1, -1, -1, -\tau^2)$. The equations of motion (Eq. (\ref{EOM})) are integrated forward in time with the help of the equation of state, ${\mathcal P} (\epsilon)$, which is obtained from lattice QCD analyses \footnote{See the discussion in Ref. \cite{Gale:2013da}, and references therein.}. The code which realizes this  is \textsc{music} \cite{Schenke:2010nt}, a 3+1D numerical hydrodynamics simulation which relies on the Kurganov-Tadmor algorithm \cite{KT}. In this work, the hydrodynamic evolution is done as in Ref. \cite{Schenke:2010nt}, with the hydro parameters specified in the last line of Table I there. The viscous evolution here uses a value of $\eta/s = 1/4 \pi$, and requires the initial energy density, $\epsilon_0$, to be 90\% of the  value in the inviscid case in order to account for entropy buildup by the dissipative dynamics.  The hydrodynamic initial state here is free of fluctuations: a quantitative study of these effects will be done in an upcoming work.

%\begin{center}
%\vspace*{.3cm}
%\begin{figure}[ht]
%\includegraphics[scale=0.32]{T_vs_tau.eps}
%\caption{(Color online) The temperature evolution as a function of the proper time, $\tau$, in a central fluid cell. The simulation with a non-zero shear viscosity coefficient uses $\eta/s = 1/4 \pi$. The parameters of the fluid dynamics simulation are given in Ref. \cite{Schenke:2010nt}.}
%\label{T_vs_tau}
%\end{figure}
%\end{center}

%It is instructive to compare the time-evolution of the bulk dynamics in the inviscid and viscous limits. This is shown in Figure \ref{T_vs_tau}, for an arbitrary  fluid cell in the transverse plane, and concentrating on the early stages. As both simulations are adjusted to reproduce the same measured final-state multiplicity, the adiabatic case needs to start with a higher temperature to make up for the lack of entropy generation. One also observes a slightly slower cooling of the viscous fluid in this time window. 

\subsection{Dilepton rates from perturbative QCD at high temperature}
\label{QGP_rates}

The production rate of dileptons depends on the local temperature. For massless quarks and antiquarks annihilating into lepton pairs, the (four-momentum integrated) rate will go as $ R \sim T^4$ \cite{Kajantie:1986dh}. Most hydrodynamical simulation models used at RHIC energies require initial and kinematical freeze-out temperatures such that the net dilepton signal should originate from temperatures both above and below the range where lattice simulations predict a transition from hadronic degrees of freedom to a phase known as the quark-gluon plasma (QGP). In the limit of zero net baryon density, lattice calculations do not show a first- or even second-order phase transition, but instead a smooth crossover centered at about 175 MeV (for a recent review see e.g. \cite{Petreczky:2012rq}). The dilepton contribution from the QGP phase will be the  most visible for invariant masses above 1 GeV, where the mass scale is large compared to both $\Lambda_{\rm QCD}$ and to $T$. For this channel, quark-antiquark annihilation at leading order (the Born approximation, for which the cross section actually doesn't depend on $\alpha_s$) is often used as an approximation to the production rate at high temperature. At lower invariant mass, processes other than the Born term will contribute. Those may be parametrically of higher order in $\alpha_s$ but typically contain collinear divergences that, when correctly re-summed, produce a final result complete to leading order in the strong coupling. Those Hard Thermal Loop (HTL) augmented rates rise over the Born result as the invariant mass is lowered, but are still only available either in the zero momentum (${\bf q} = 0$ \cite{Braaten:1990wp}) or large energy ($q^0 \gtrsim T$ \cite{Aurenche:2002wq}) limits. Recently, the vector current correlation function has also been evaluated on the lattice, enabling the extraction of a thermal dilepton production rate at a given temperature and vanishing pair three-momentum \cite{Ding:2010ga}. The Born rate is used in this work; in part because of the kinematical restrictions still associated with the newer dilepton production rates, but mostly because the rates from the confined hadronic sector of QCD will produce a dominant contribution at the invariant masses considered here (and for conditions prevalent at RHIC) \cite{Rapp:2013nxa}. In addition, this formulation is readily amenable to a viscous correction, as will be discussed in the next section. 

We start with high temperatures and use perturbative QCD. The cross-section for $q \bar{q} \to \ell^+ \ell^-$ is, neglecting quark and lepton masses, 
%
%\begin{figure}[!ht]
%\begin{center}
%\includegraphics[scale=0.7]{QGP.eps}
%\end{center}
%\caption{Feynman diagram used in the calculation of the Born QGP rate.}\label{pic:QGP_rate}
%\end{figure}
%
%
\begin{equation}
\sigma = \frac{16\pi \alpha_{\rm EM}^2 \left(\sum_{q'} e^2_{q'}\right) N_c}{3 q^2} 
\end{equation}
where the index $q'$ runs over quark flavours, and $q$ is the four-momentum of the virtual photon. The rate of dilepton production in the Born approximation is related to the cross-section through 
\begin{eqnarray}
\frac{d^4 R}{d^4 q} = \int \frac{d^3 p_1 d^3 p_2}{(2\pi)^6 p^0_1 p^0_2 } n_F(p_1) n_F(p_2) \frac{q^2}{2} \sigma \delta^4(q-p_1-p_2) 
\end{eqnarray}
where $p_1$ and $p_2$ label the momenta of the incoming quarks and $n_F$ is the Fermi-Dirac distribution. Integrating gives  
\begin{eqnarray}
\frac{d^4 R}{d^4 q} = \frac{\alpha_{\rm EM}^2}{6\pi^4} \frac{1}{\exp(\beta q^0)-1}\left\{1-\frac{2}{\beta|{\bf q}|} \ln \left[ \frac{n_{-}}{n_{+}}\right] \right\}
\end{eqnarray} 
where $n_{-}=1+\exp\left[ \frac{-\beta(q^0-|\bf q|)}{2}\right]$, $n_{+}=1+\exp\left[ \frac{-\beta(q^0+|\bf q|)}{2}\right]$, $\beta=1/T$, $|\bf q|$ is the norm of the three-momentum of the virtual photon, $q^0$ is its energy, and we used $q^{\prime}=u,d,s$.

When gradients of the flow or the temperature exist in a viscous fluid, departures from thermal equilibrium must occur, and those will modify the distribution functions.  These departures should be reflected in the traces taken over the density matrix to determine what were previously thermal averages: $n_{\rm eq} (p) \to n_{\rm eq} (p)  + \delta n (p)$. Various ans\"atze for the form of $\delta n(p)$, were examined in \cite{dusling-moore-teaney}. In this work we use  
\begin{equation}
\delta n(p) = \frac{C}{2}n(p)(1 \pm n(p))\frac{p^\alpha p^\beta}{T^2} \frac{\pi_{\alpha \beta}}{\epsilon+P} 
\label{delta_n}
\end{equation}
where $C$ is at this point an undetermined proportionality constant. Calculating $T^{\mu \nu}$ with this expression and matching this to $T^{\mu \nu}_{ \rm ideal} + \pi^{\mu \nu}$ (defined in the previous section)  gives $C_q = \frac{7\pi^4}{675\zeta(5)}\approx 0.97$ for the Fermi-Dirac distribution of a single component massless quark fluid. %Both the energy-momentum tensor and its viscous correction are calculated using \textsc{music}. 

We modify the thermal average with this change in the Fermi-Dirac distribution where it appears, as was done in \cite{Dusling:2008xj}. The details of this derivation are presented in Appendix \ref{QGPViscousCorrection}. This yields a viscous correction to dilepton production dependent on $\pi^{\mu \nu}$:
\begin{eqnarray}
\frac{d^4 R}{d^4 q}       & = & \frac{d^4 R_0}{d^4 q} + \frac{d^4 \delta R}{d^4 q}\nonumber\\
\frac{d^4 R_0}{d^4 q}        & = & \frac{q^2}{2} \frac{\sigma}{(2\pi)^5}\frac{1}{\exp(\beta q^0)-1}\left\{1-\frac{2}{\beta |\bf q|} \ln \left[ \frac{n_{-}}{n_{+}}\right] \right\} \nonumber\\
\frac{d^4 \delta R}{d^4 q} & = & \frac{q^2}{2} \frac{\sigma}{(2\pi)^5} C_q \frac{q^\alpha q^\beta}{T^2} \frac{\pi_{\alpha\beta}}{\epsilon+P} \frac{1}{2|{\bf q}|^5} \int dE_1 n(E_1)n(q^0-E_1)(1-n(E_1)) D \nonumber\\
                       D  & = & \left[(3 q_0^2 -|{\bf q}|^2) E^2_1-3 q^0 E_1 q^2 + \frac{3}{4} q^4\right]
\end{eqnarray}
The rates now depend on accurate calculation of the non-ideal corrections to the energy-momentum tensor, and therefore require viscous hydrodynamical simulation at RHIC and the LHC, such as \textsc{music}. In Section \ref{production}, the implications of the viscous correction on observables are explored.

%%%%%%%%%%%%%%%%%%%%%%%%%%%%%%%%%%%%%%%%%%%%%%%%%%%%%%%%%%%%%%%%%%%%%%%%%%%
\subsection{Rates from a hadronic medium }\label{hg_rates}
%%%%%%%%%%%%%%%%%%%%%%%%%%%%%%%%%%%%%%%%%%%%%%%%%%%%%%%%%%%%%%%%%%%%%%%%%%%%

Most of the thermal production of dileptons from a hadronic medium  are produced by low mass vector mesons $V$ ($V=\rho, \; \omega, \; \phi$). In fact the majority of the the emission below $M<1.2$ GeV is of hadronic origin. A description for this production requires an accurate effective coupling between the electromagnetic field and the hadrons, which are composite particles and have, in general, complicated and mostly unknown electromagnetic form factors. The vector meson dominance model (VMD), first proposed by Sakurai \cite{gounaris-sakurai}, successfully describes dilepton production \cite{kapusta-gale-book,rapp}. The effective coupling in this model is given by
\begin{equation}
\mathcal{L} = \mathcal{L}_{QED} - \sum_{V = \rho, \omega, \phi} \left[\frac{\sqrt{4\pi\alpha_{\rm EM}}}{g_V}m^2_{V}V^{\mu}A_{\mu} + \frac{1}{4} F^{\mu \nu}_V F^V_{\mu \nu} \right]
\end{equation}
where where $F^{\mu\nu}_V=(\partial^\mu V^\nu-\partial^\nu V^\mu)$, and  $\mathcal{L}_{QED}= \bar{\psi}_{\ell} \left( i\not \! \partial - m_{\ell}\right) \psi_{\ell} - \sqrt{4\pi\alpha_{\rm EM}}\, \bar{\psi}_{\ell}\gamma^{\mu}\psi_{\ell} A_{\mu}- \frac{1}{4}F^{\mu \nu}F_{\mu \nu}$ . The coupling constants $g_V$ are determined by measuring the vacuum decay rate of vector mesons to dileptons. The thermal rate of dilepton production for each low mass vector meson $V$ is then
\begin{eqnarray}
\frac{d^4 R_V}{d^4 q}= - \frac{\alpha_{\rm EM}^2}{\pi^3}\frac{L(M)}{M^2}\frac{m^4_V}{g^2_V}\left[\frac{{\rm Im}\, D^{\rm R}_V}{e^{\beta q^0}-1}\right] \label{eq:HG_rate}
\label{idealHGRate}
\end{eqnarray}
where $L(M)=\left(1+\frac{2m^2_\ell}{M^2}\right)\sqrt{1-\frac{4m^2_\ell}{M^2}}$, and the imaginary part of the retarded vector propagator, ${\rm Im}\,D_V^{\rm R} = \frac{1}{3}{\rm Im} D^{\mu {\rm R}}_\mu$, was obtained  using the Kubo-Martin-Schwinger (KMS) relation and VMD \cite{kapusta-gale-book}. From here on, we shall set the lepton mass $m_{\ell}$ to zero.

Viscous effects will modify the retarded self-energy and any averages that were originally thermal. To estimate these changes, we will need to determine the viscous corrections to the thermal emission rate, Eq.(\ref{idealHGRate}). In a kinetic theory formulation of the dilepton production, the microscopic equilibrium distribution functions can be replaced with their non-equilibrium counterparts, and the net rates can then be recalculated in the different hadronic channels, as was done for real photons \cite{dion-paquet-schenke-young-jeon-gale}, and for the Born contribution of the previous section. Note that an  empirically successful modelling of the electromagnetic current-current correlator for an interacting ensemble of baryons and mesons is achieved through hadronic many-body theory \cite{Rapp:1999ej,*Rapp:2009yu}, and builds on Eq. (\ref{idealHGRate}). At this point, the viscous corrections to dilepton rates resulting from this approach have yet to be derived, and the non-perturbative extension of the KMS relation to the non-equilibrium realm is still a topic in development. Therefore in this work a formalism based on experimental data is adopted, and the consequences of shear viscous corrections to the self-energy are explored as detailed below.
 
The total vector meson self-energy is given by \cite{eletsky-belkacem-ellis-kapusta}:
\begin{equation}
\Pi^{\rm tot}_{V} (M, |{\bf p}|, T)= \Pi^{\rm vac}_V\left( M \right) +  \Pi^{\rm T}_V \left(M, |{\bf p}|, T\right) +  \delta\Pi^{\rm T}_V\left(M, |{\bf p}|, T \right)
\label{SelfE}
\end{equation}
The vacuum self-energy depends only on invariant mass $M=\sqrt{E^2-|{\bf p}|^2}$, while the finite temperature contributions depend on $E$ and $|{\bf p}|$ (or $M$ and $|{\bf p}|$) as do the viscous corrections to the self-energy. Here as in \cite{eletsky-belkacem-ellis-kapusta}, $\Pi^{\rm T}_V$ is evaluated on the mass shell of $V$. 

The calculations of $\Pi^{\rm vac}_V$ are present in \cite{eletsky-belkacem-ellis-kapusta,martell-ellis,vujanovic-ruppert-gale}; there, terms in the Lagrangian describe all interactions contributing to $\Pi^{\rm vac}_V$. We make Eq. (\ref{SelfE})  more explicit by writing $\Pi^{\rm T}_{Va}$ and relating it to the  vacuum forward scattering amplitude $f_{Va}(s)$ as in \cite{eletsky-belkacem-ellis-kapusta, jeon-ellis}:
\begin{equation}
\Pi^{\rm T}_{Va} = -4 \pi \int \frac{d^3k}{(2\pi)^3} n_a(u\cdot k) \frac{\sqrt{s}}{\omega}f_{Va}(s)
\label{eq:thermal_self-energy}
\end{equation}	
where $V$ represents the vector meson and $a$ the particle with which the vector meson interacts. Also, $n_a$ is a Fermi-Dirac or Bose-Einstein distribution function of the particle of type $a$, $k^\mu$ is its four-momentum, while $u^\mu$ is the velocity of the fluid cell. The work by Eletsky {\it et al.} \cite{eletsky-belkacem-ellis-kapusta} describes how to obtain the thermal correction to the vacuum self-energy of the vector meson $V$ using $f_{Va}(s)$ and Eq.(\ref{eq:thermal_self-energy}). For a more in-depth derivation of the viscous correction to the self-energy, the reader is refered to Appendix \ref{HGViscousCorrection}. 
%This expression assumes that the transverse and longitudinal components of the in-medium self-energy are the same (it is often the case that they are very similar), that the energy and the momentum of a particle in the medium are still related by the vacuum relation $\omega=\sqrt{{\bf k}^2+m^2_a}$, and that the density of the medium is low so that there are no coherence effects between two consecutive scatterings.

Using the relation between the self-energy and $f_{Va}$, and modifying the thermal average with viscous corrections, gives an expression for $\delta \Pi^{\rm T}_{Va}$ that is dependent on the non-ideal correction to $T^{\mu \nu}$. Quoting from Appendix \ref{HGViscousCorrection}, the result is 
\begin{eqnarray}
\label{rest_frame}
\Pi^{\rm tot}_{V} (M, |{\bf p}|) & = & \Pi^{\rm vac}_V\left( M \right) \\
                                 & - & \sum_{a=N,\bar{N},\pi} \frac{m_V m_a T}{\pi |{\bf p}|}\int^{\infty}_{m_a} d\omega'\ln\left[\frac{1\pm\exp\left(-\omega_{+}/T\right)}{1\pm\exp \left(-\omega_{-}/T\right)}\right]f^{\rm a's\text{ }rest}_{Va}\left(\frac{m_V}{m_a} \omega' \right) \nonumber\\
                                 & + & \sum_{a=N,\bar{N},\pi} C_a B_{2,Va} \frac{p^{\alpha}_V p^{\beta}_V}{T^2} \frac{\pi_{\alpha\beta}}{\epsilon+P} \nonumber 
\end{eqnarray}
where Eq.(\ref{B2}) gives an expression for $B_{2,Va}$. Unlike the case of the QGP where considering a single component fluid was sufficient to compute $C_q$, the hadronic medium  is a mixture of many particle species. A simplifying assumption was made: $C_a$ is particle independent \cite{monnai-hirano}, and $\forall a: C_a=1$. Also, this work is done in the limit where the self-energies arising from interactions with anti-nucleons and nucleons are the same. 

Physics relevant to interacting hadrons now enters this expression through $f_{Va}$, which receives contributions at both low and high energies; $f_{Va}$ at low energy has both resonance and pomeron contributions. In the center-of-mass (c.m.) frame \cite{eletsky-belkacem-ellis-kapusta}
\begin{equation}
f^{\rm c.m.}_{V a}(s) = \frac{1}{2q_{\rm c.m.}}\sum_{R} W^{R}_{Va}\frac{\Gamma_{R\rightarrow Va}}{M_{R} - \sqrt{s} - \frac{1}{2}i\Gamma_{R}} - \frac{q_{\rm cm}}{4 \pi s} \frac{1+\exp(-i \pi \alpha_P)}{\sin(\pi \alpha_P)} r_{Va}^{P} s^{\alpha_P} \label{eq:f_low}
\end{equation}
The center of mass momentum is $q_{\rm c.m.}$, which can be expressed in terms of the Mandelstam variable $s$. Here the sum ranges over resonances $R$ that decay into the vector meson $V$ and the particle $a$, which is either a nucleon or a pion. The spin and isospin are averaged, leading to the factor $W^{R}_{Va} = \frac{(2s_R + 1)}{(2s_{V}+1)(2s_a+1)} \frac{(2t_R + 1)}{(2t_{V} + 1)(2t_a + 1)}$, with $s_i$ being the spin of particle $i$, and $t_i$, its isospin. $\Gamma_{R\rightarrow Va}$ is an effective width of the decay of the resonance $R$ into $Va$. Its internal structure, and the types of resonances contributing to $f_{Va}$, are all discussed in detail in \cite{eletsky-belkacem-ellis-kapusta,martell-ellis,vujanovic-ruppert-gale}. The values of the Regge residues $r_{Va}$, intercept $\alpha_P$, and the resonances included in $f_{Va}$  are all given there.  The transformation of the distribution function from the rest frame of $a$ (as used in Eq. (\ref{rest_frame})) to the c.m. frame of particles $V$ and $a$ (Eq. (\ref{eq:f_low})) is straightforward \cite{kapusta-gale-book}. 

The high-energy limit of $f_{Va}$ is described by a Regge parametrisation \cite{eletsky-belkacem-ellis-kapusta,martell-ellis,vujanovic-ruppert-gale}:
\begin{equation}
f^{\rm c.m.}_{V a} \left(s\right)= -\frac{q_{\rm c.m.}}{4 \pi s} \sum_{i} \left[ \frac{1+\exp(-i \pi \alpha_i)}{\sin(\pi \alpha_i)} \right] r_{Va}^{i} s^{\alpha_i} \label{eq:f_high}
\end{equation}
The low-energy and the high-energy pieces are then matched onto one another at $E_V-m_V \sim 4$ GeV for pions and $E_V-m_V \sim 1$ GeV for nucleons, where $E_V$ is evaluated in the rest frame of pions and nucleons respectively. To verify that the matching does not introduce violations of the Kramers-Kronig relations, a dispersion integral formula relating the real part of $f_{Va}$ to a principal value integral over its imaginary part \cite{eletsky-belkacem-ellis-kapusta,vujanovic-ruppert-gale} is used:
\begin{equation}
{\rm Re}\left[f_{V a} \left(E_V \right) \right] = {\rm Re} \left[f_{V a} \left( 0 \right) \right] + \frac{2E^{2}_{V}}{\pi} \mathrm{P.V.} \int^{\infty}_{m_V} \frac{{\rm Im} \left[ f_{Va}\left( E^{\prime}\right) \right]  d E^{\prime}}{E^{\prime}\left( E^{\prime}+E_{V}\right) \left( E^{\prime} - E_{V}\right)}
\end{equation}
Indeed, as was shown in \cite{eletsky-belkacem-ellis-kapusta,martell-ellis,vujanovic-ruppert-gale} the effect of the matching procedure on the shape of the forward scattering amplitude is not significant.  The dilepton rates derived in Sections \ref{QGP_rates} and \ref{hg_rates} have been used previously in an interpretation of NA60 data \cite{Ruppert:2007cr}, taken at the CERN SPS.

%%%%%%%%%%%%%%%%%%%%%%%%%%%%%%%%%%%%%%%%%%%%%%%%%%%%%%%%%%%%%%%%%%%%%%%%%%%%%%%%%%%%%%%%%%%%%%%%%%%%%%%%%%%%%%%%%%%%%%%%%%%%%%%%%%%%%%%%%%%%%%%%
\section{Lepton pairs from charm decays}%%%%%%%%%%%%%%%%%%%%%%%%%%%%%%%%%%%%%%%%%%%%%%%%%%%%%%%%%%%%%%%%%%%%%%%%%%%%%%%%%%%%%%%%%%%%%%%%%
\label{charm}%%%%%%%%%%%%%%%%%%%%%%%%%%%%%%%%%%%%%%%%%%%%%%%%%%%%%%%%%%%%%%%%%%%%%%%%%%%%%%%%%%%%%%%%%%%%%%%%%%%%%%%%%%%%%%%%%%%%%%%%%%%%%%%%%%%
%%%%%%%%%%%%%%%%%%%%%%%%%%%%%%%%%%%%%%%%%%%%%%%%%%%%%%%%%%%%%%%%%%%%%%%%%%%%%%%%%%%%%%%%%%%%%%%%%%%%%%%%%%%%%%%%%%%%%%%%%%%%%%%%%%%%%%%%%%%%%%%%

Dileptons originate not only from electromagnetic transitions, but also from weak decays: a charm quark decays semi-leptonically into an electron with a branching fraction of approximately 10\%. In a proton-proton collisions with center-of-mass energies of 200 GeV, dileptons produced from pairs of charm quarks dominate the yield in the intermediate mass range (from 1.2 GeV to 2.5 GeV) \cite{Zhao:2011wa}. Therefore, the analysis of the dilepton spectrum provides a measurement of the charm cross section. In this work however, the emphasis is placed on the interaction (energy loss and gain, angular deflection)  of heavy quarks  with the hot and dense viscous matter, and how this  will reflect itself in the dilepton spectrum. The production of heavy quarks in relativistic nuclear collisions - and their interaction with the hot and dense medium - is a topic that has received much attention over the recent years \cite{Shuryak:1996gc,vanHees:2005wb,*PhysRevLett.100.052301,*Gossiaux:2012th,*Younus:2013rja,Moore:2004tg}. 

The mass of a charm quark pair is much greater than the temperature reached in any model of the heavy-ion collisions at RHIC or the LHC; thermal production is negligible in comparison with the partonic annihilation in the initial collision. The mass of a charm quark pair is also significantly larger than $\Lambda_{\rm QCD}$, and the production can be treated perturbatively. For proton-proton collisions, fixed-order next-to-leading-log (FONLL) calculations \cite{Cacciari:1998it} fit the available experimental data well by including both next-to-leading order results at low momenta and terms proportional to $\alpha_s \log(p/m)$ and $\alpha_s^2(\log(p/m))^2$, and by treating the heavy quarks as effectively massless at large $p_T$. In heavy-ion collisions, the initial production of charm (and anti-charm) is affected by changes in the parton distribution functions: there can be - depending on the energy scale - shadowing/anti-shadowing  of the parton distribution functions as well as isospin dependence of the heavy quark cross sections. The measured nuclear parton distribution functions can be evolved to different values of $Q$ with the DGLAP equations. Then, one needs to calculate the effect of the in-medium evolution of heavy quarks in heavy-ion collisions. The transport coefficients for heavy quarks have proven to be difficult to estimate reliably with hard-thermal-loop effective theory \cite{caron-huot}; however, for heavy quark momenta both less than and on the order of the heavy quark mass, the evolution of heavy quarks can be approximated to be diffusive and relativistic Langevin equations describe their dynamics \cite{Moore:2004tg}, allowing the heavy quark diffusion coefficient to be estimated phenomenologically. 

We use \textsc{pythia8} to generate events with heavy quarks.  We also use \textsc{eks98} to determine the initial parton distribution functions in the nuclei. Then, using the same hydrodynamical description as was used to determine the thermal dilepton production, the heavy quarks are evolved using relativistic Langevin dynamics and the heavy quark spatial diffusion coefficient $D_c = 3/(2\pi T)$. The heavy quarks then hadronize according to Peterson fragmentation \cite{peterson} into $D$, $\bar{D}$, $D^*$, and $\Lambda_c$ particles that then decay semi-leptonically. The quantitative results of our modelling are reported in  Section \ref{charmresults}.

%In-medium evolution of heavy quarks has a significant impact on dileptons in the intermediate mass range. First of all, energy loss leads to significant depletion of $dM/dy$ in this range. Second of all, the flow of heavy quarks leads to azimuthal anisotropy of the dileptons. These effects will be explored quantitatively in Section \ref{charmresults}.

%%%%%%%%%%%%%%%%%%%%%%%%%%%%%%%%%%%%%%%%%%%%%%%%%%%%%%%%%%%%%%%%%%%%%%%%%%%%%%%%%%%%%%%%%%%%%%%%%%%%%%%%%%%%%%%%%%%%%%%%%%%%%%%%%%%%%%%%%%%%%%%%
\section{Results }%%%%%%%%%%%%%%%%%%%%%%%%%%%%%%%%%%%%%%%%%%%%%%%%%%%%%%%%%%%%%%%%%%%%%%%%%%%%%%%%%%%%%%%%%%%%%
\label{production}%%%%%%%%%%%%%%%%%%%%%%%%%%%%%%%%%%%%%%%%%%%%%%%%%%%%%%%%%%%%%%%%%%%%%%%%%%%%%%%%%%%%%%%%%%%%%%%%%%%%%%%%%%%%%%%%%%%%%%%%%%%%%%
%%%%%%%%%%%%%%%%%%%%%%%%%%%%%%%%%%%%%%%%%%%%%%%%%%%%%%%%%%%%%%%%%%%%%%%%%%%%%%%%%%%%%%%%%%%%%%%%%%%%%%%%%%%%%%%%%%%%%%%%%%%%%%%%%%%%%%%%%%%%%%%%
\subsection{Thermal dilepton yield: the transverse momentum and invariant mass dependence}
%\subsubsection{Impact of the viscous corrections $\delta n$ of the HG rates on the yield}

\begin{figure}[!ht]
\vspace{0.35cm}
\begin{center}
\includegraphics[scale=0.3]{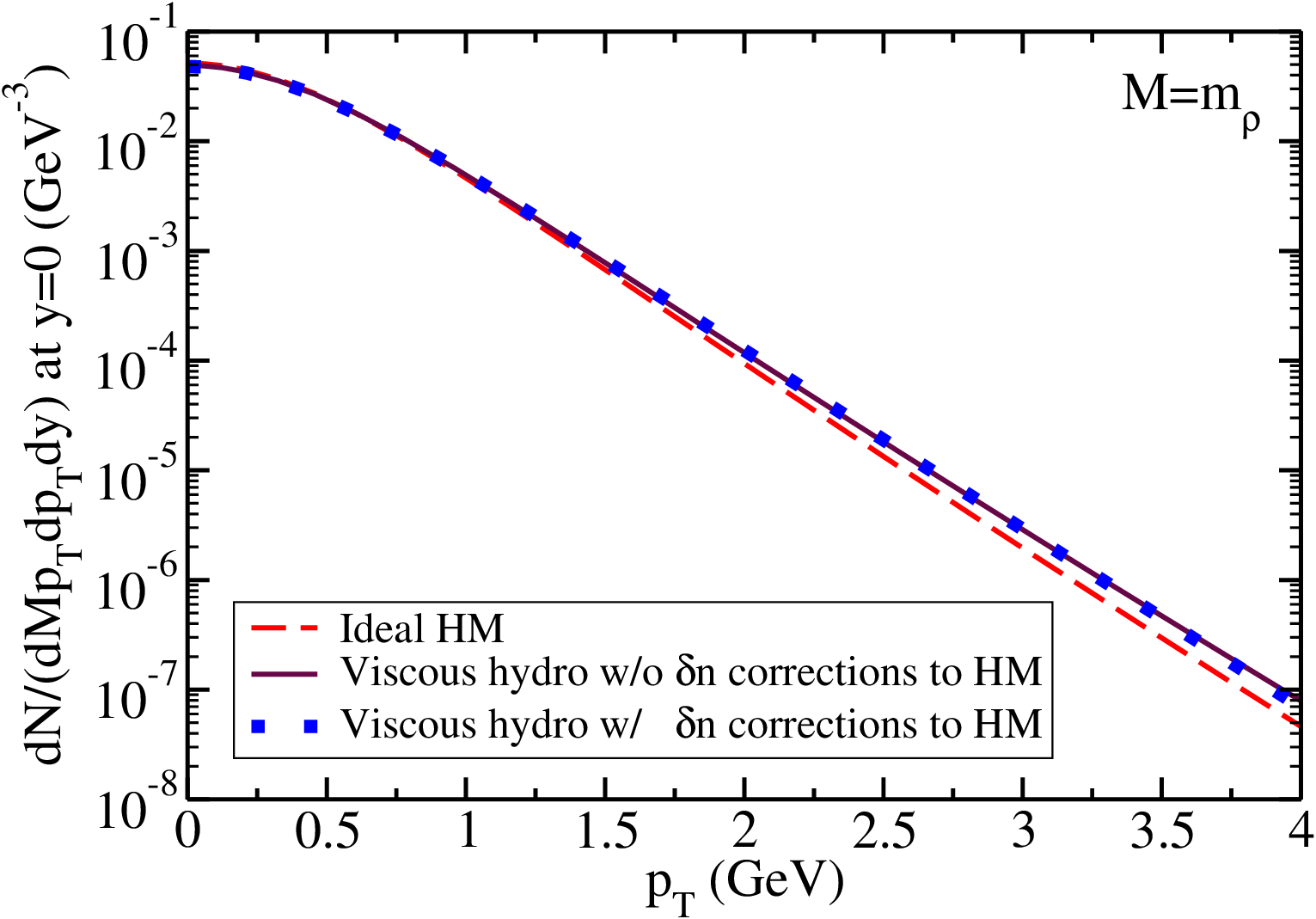} \hspace*{1cm}  \includegraphics[scale=0.3]{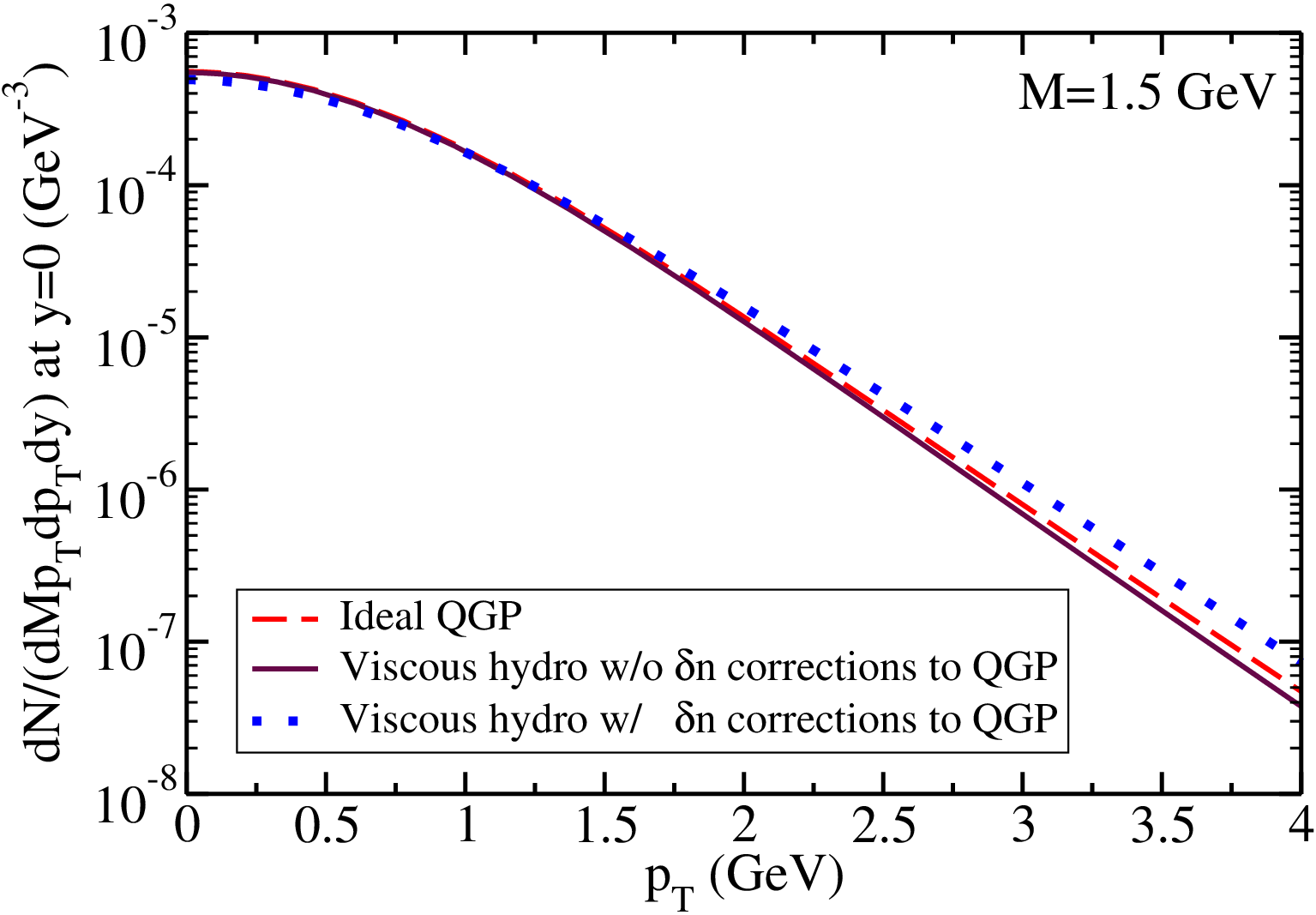}
\end{center}
\caption{(Color online) Left panel: Dilepton yield from the hadronic medium (HM) only, in the 0-10\% centrality class and fixed invariant mass $M=m_\rho$. The contribution from: (i) ideal hydro evolution (dashed line), (ii) viscous hydro evolution alone (solid line), and (iii) viscous hydro evolution including viscous corrections to ideal dilepton rates are shown (square dots). Right panel: Dilepton yield from the QGP only, in the same centrality class, and for $M=$ 1.5 GeV.}
\label{pic:yield_ideal_visc_w_wo_df}
\end{figure}

The yield of lepton pairs is obtained in our approach by integrating the production rates over the space-time history of the collision, using relativistic hydrodynamics to simulate the time- and  space-dependent background fields. It is  instructive to compare the transverse momentum spectra associated with different values of the dilepton invariant mass. In order to highlight in turn the hadronic and QGP thermal contributions, two values chosen can be associated with the ``low mass region'' ($M = m_\rho$), and the ``intermediate mass region'' ($M =$ 1.5 GeV), respectively. We first consider the effect of viscous corrections only on the dileptons originating from the hadronic matter phase. 
In Fig. \ref{pic:yield_ideal_visc_w_wo_df} (left panel), the dilepton yields as a function of $p_T$ for the 0-10\% centrality at a fixed invariant mass $M=m_\rho$ are plotted, considering in turn three cases: that of inviscid hydrodynamics, then allowing for viscous corrections to the bulk evolution but not to the rates, and then finally correcting both the rates and the bulk evolution. 
\begin{figure}[!bt]
\vspace{.3cm}
\begin{center}
\includegraphics[scale=0.305]{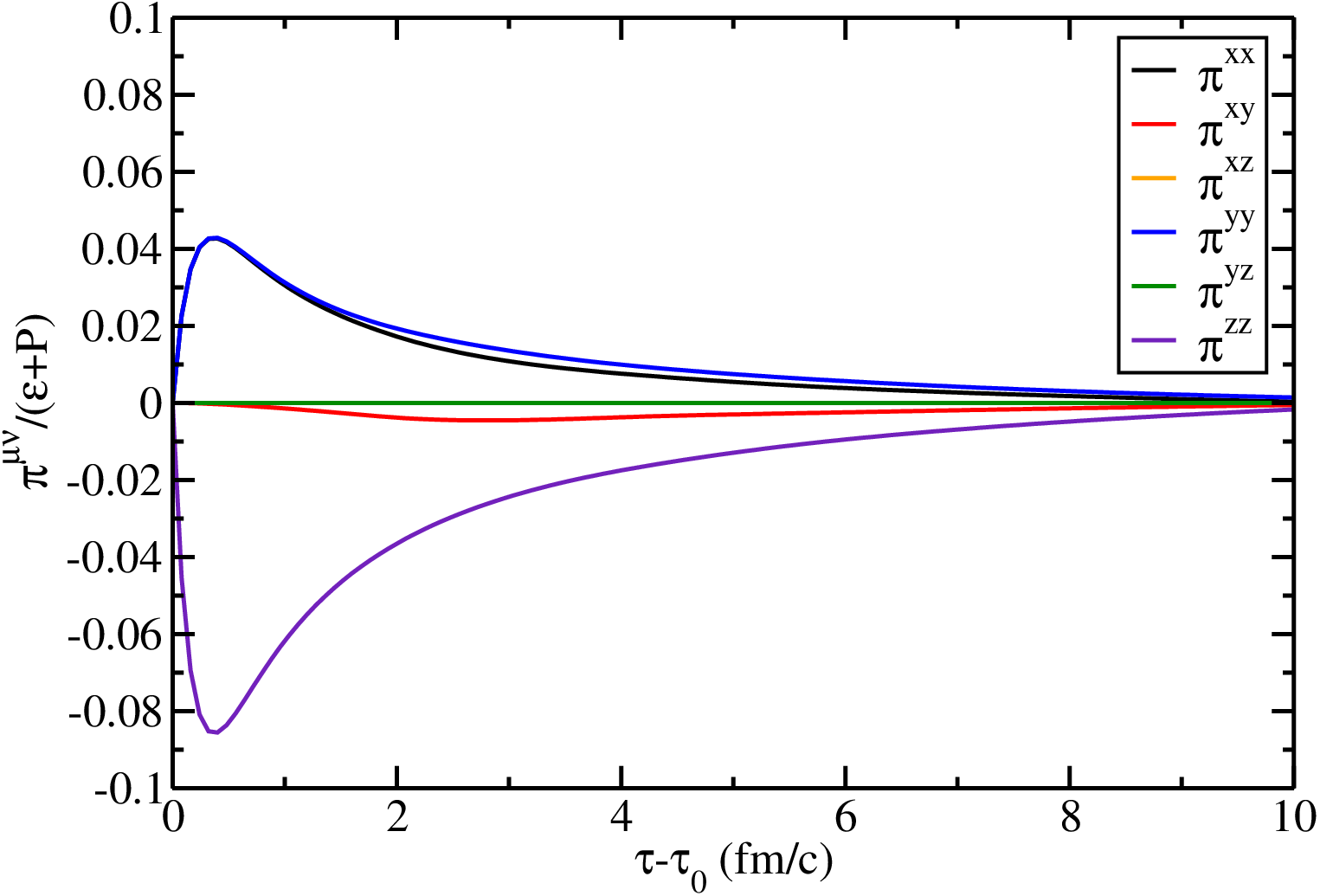}\hspace*{1cm}\includegraphics[scale=0.3]{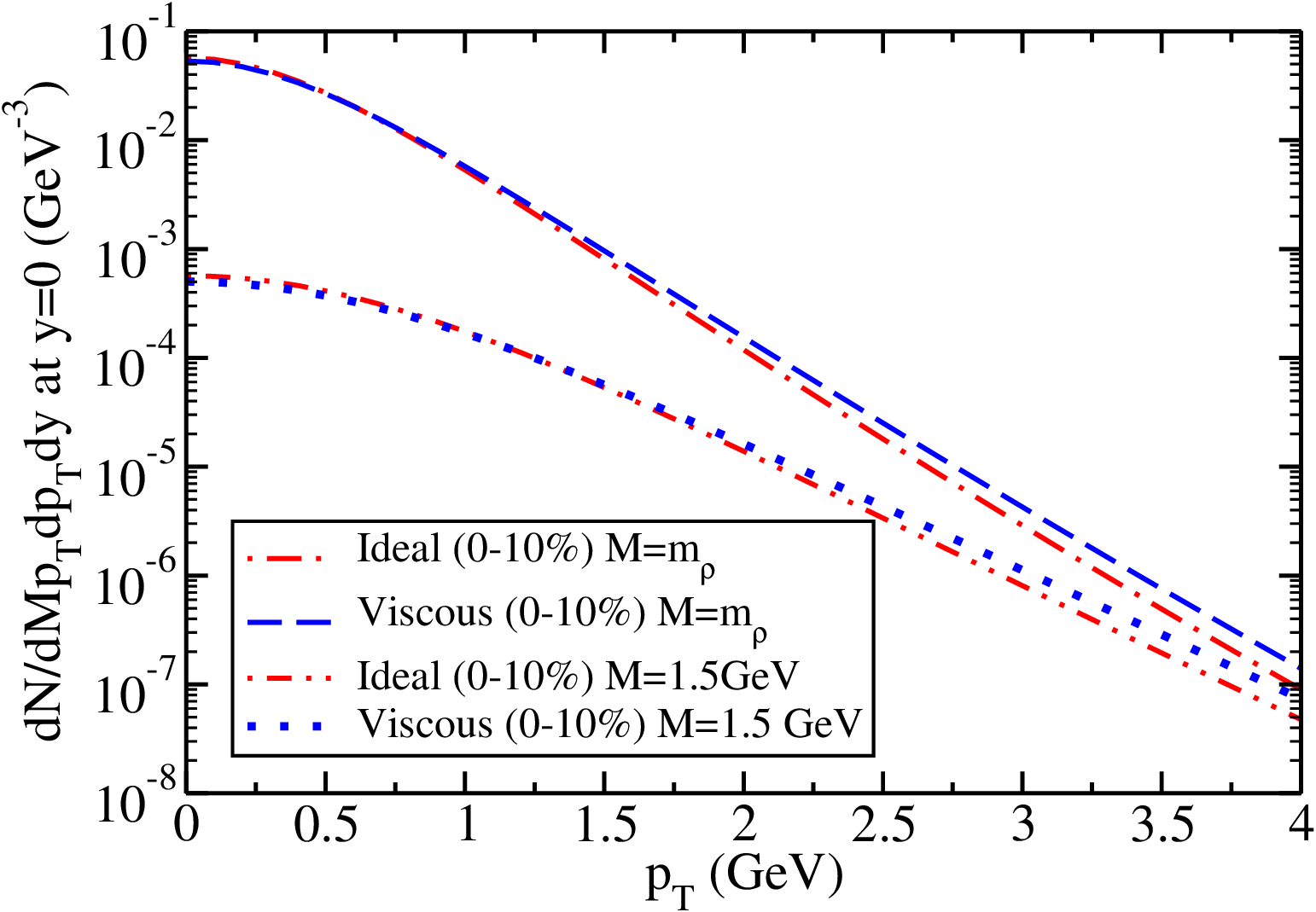}
\end{center}
\caption{(Color online) Left panel: Shear stress tensor in the local rest frame of the cell located at $x = y =$ 8/3 fm $z =$ 0 fm in the 0-10\% centrality class. Right panel: Total thermal dilepton yield (HM+QGP) as a function of $p_T$ and at two different invariant masses: $M= m_\rho$ and $M = 1.5$ GeV.}
\label{pic:pi_munu}
\end{figure} 
The viscous effects on the bulk evolution in the hadronic phase raise the yield slightly ($\sim 60\%$) at momenta from 3 to 4 GeV, as the viscous evolution slows down the temperature drop in the high-$T$ portion of the hadronic phase \cite{Song:2007ux,dion-paquet-schenke-young-jeon-gale}. 
We also notice, on the scale of the plot in the left panel of Fig. \ref{pic:yield_ideal_visc_w_wo_df}, that viscous corrections to the hadronic emission rates have basically no effect over that of the viscous evolution. The physical reason explaining the irrelevance of $\delta n$ corrections on the yield arises from the fact that dileptons from the hadronic phase are mostly emitted late ($\tau\gtrsim4$ fm/c) at which time $\pi^{\mu\nu}$ is small (see the left panel of Fig. \ref{pic:pi_munu}). Note that this explanation is somewhat qualitative as many cells with different temperatures contribute to the net dilepton yield. But statement is verified by a direct calculation, and viscous photon yields  exhibit the same behaviour \cite{dion-paquet-schenke-young-jeon-gale}.   Turning to dileptons from the QGP phase only, this receives further support from the dilepton transverse momentum spectrum for $M=$ 1.5 GeV, shown on the right panel of Fig. \ref{pic:yield_ideal_visc_w_wo_df}. Correcting the bulk evolution only leads to a slight decrease of the yield at transverse momentum values of $p_T \sim 2-4$ GeV. This is because the initial temperature in the viscous case is lower than that in the inviscid case, owing to entropy generation \footnote{See, for example, Figure 1 in Ref. \protect\cite{dion-paquet-schenke-young-jeon-gale}}: recall that the entropy in the final state is directly related to the observed particle multiplicity. Unlike the case of the hadron medium, the $\delta n$ correction does influence the net dilepton yield as the emission occurs at early times when the temperature is high, which coincides with the proper time interval where the magnitude of the shear pressure tensor is maximal. 
%%%%%%%%%%%%%%%%%%%%%%%%%%%%%%%%%%%%%%%%%%%%%%%%%%%%%%%%%%%%%%%%%%%%%%%%%%%%
%\subsubsection{Dilepton yields at low invariant mass}
%%%%%%%%%%%%%%%%%%%%%%%%%%%%%%%%%%%%%%%%%%%%%%%%%%%%%%%%%%%%%%%%%%%%%%%%%%%%
%

The right panel of Figure \ref{pic:pi_munu} displays the net thermal dilepton yield (includes both HM and QGP contributions) as a function of transverse momentum in the 0 - 10\% centrality class, for two values of invariant mass. For invariant masses in the low mass region, the higher momentum yield's sensitivity to the shear viscosity coefficient manifests itself almost exclusively through that of the bulk evolution. On the other hand, the thermal yield at higher invariant masses shows that the initial conditions (here, mainly $T_i$, the initial hydro temperature), the hydro evolution, and the viscous corrections to the distribution functions all have an effect.   While the different ingredients invoked here leave a quantitative imprint on the dilepton transverse momentum spectrum that is still quantitatively modest, these findings do confirm the power and the potential of lepton pairs as both a precise thermometer and viscometer. We leave the search for the specific conditions (e. g. centrality classes, different initial state conditions, beam energy scans, etc.) that will accentuate and perhaps even maximize those differences to an upcoming study.

%We start by showing our results for the dilepton yields as a function of $p_T$ (see Fig. \ref{pic:yield_ideal_vs_visc_pt_hg_qgp}), including all viscous $\delta n$ corrections. 
%
%\begin{figure}[!ht]
%\vspace{0.85cm}
%\begin{center}
%\includegraphics[scale=0.35]{Ideal_vs_Viscous_pt_yield_improved.eps}
%\end{center}
%\caption{(Color online) Dilepton yield from hadronic gas and QGP at minimum bias and fixed invariant mass $M=m_\rho$, the dilepton yield from hadronic gas  coming from ideal hydrodynamical evolution (solid line), and fully viscous evolution (dashed line). For the QGP the ideal yield (dashed-dotted line) and fully viscous yield (dashed-double-dotted line) are illustrated.}\label{pic:yield_ideal_vs_visc_pt_hg_qgp}
%\end{figure} 
%
%Notice that the ``fully viscous'' yields are typically higher than ideal yields, and that difference widens as momentum increases reaching about 60\% at $p_T\sim4$ GeV. This is a consequence of viscosity slowing down the expansion rate of the medium and hence the medium is not able to cool as fast as in the ideal case. This is a result of our initialization $\pi^{\mu\nu}=0$. As noted in the previous section, this 60\% difference is not easily measurable. 

\begin{figure}[!h]
\vspace{0.60cm}
\begin{center}
\includegraphics[scale=0.35]{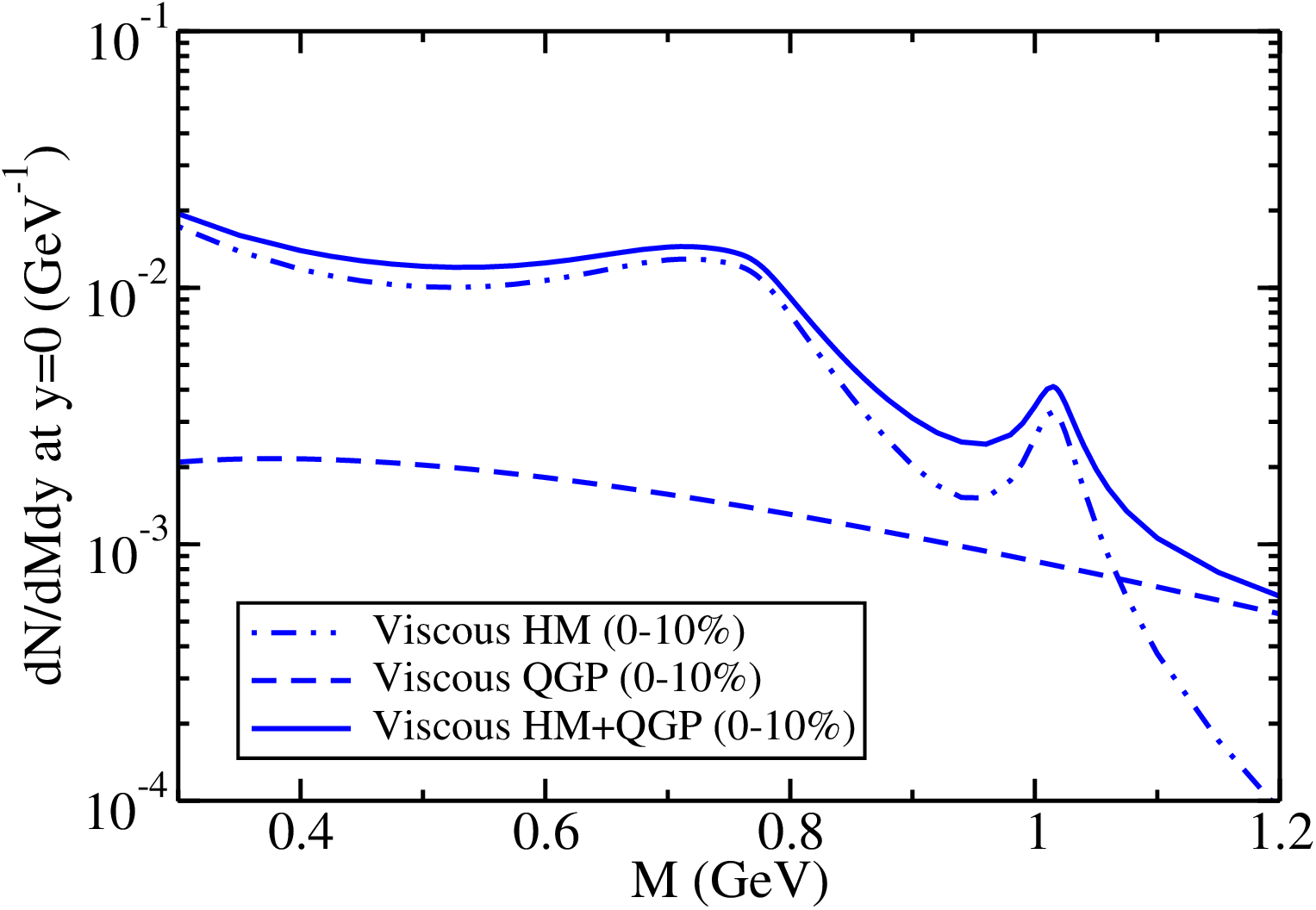}
\end{center}
\caption{(Color online) Dilepton yield from hadronic medium  and QGP  as a function of invariant mass, in the 0 - 10\% centrality class. }\label{pic:yield_ideal_vs_visc_M_hg_qgp}
\end{figure}

The effect of viscous corrections to the dilepton invariant mass distribution is now investigated. It is straightforward to show that, owing to defining symmetry properties of the shear pressure tensor ($u_\mu \pi^{\mu \nu} = u_\nu \pi^{\mu \nu} = 0$ in the fluid rest frame and  $\pi^\mu_\mu = 0$), the viscous corrections to the QGP and HM dilepton rates as a function of the invariant mass $M$ vanish: $d \, \delta R / dM = 0$. Hence, the differences between the invariant mass profiles in the inviscid and viscous cases entirely stem from the different time-evolutions. For the conditions in this study, those appropriate for RHIC, the viscous evolution has an effect on the thermal dilepton spectrum that is essentially indistinguishable for that of the ideal hydrodynamic evolution: only the viscous case is plotted in Figure \ref{pic:yield_ideal_vs_visc_M_hg_qgp}. The dilepton yield itself is therefore a poor viscometer. The spectrum asymmetry | as quantified by the elliptic flow | is now investigated. 
%
%Fig. \ref{pic:yield_ideal_vs_visc_M_hg_qgp} shows how $dN/dM$ for dileptons is insensitive to the presence of viscous effects (caused by either the modified hydrodynamical evolution or the correction to $\delta n$). Since the curves are superimposed, we decided to plot solely the viscous case. In both situations, the normalization of the yield is identical thanks to the fact that we lowered initial energy density in viscous hydrodynamics compared to the density used for ideal hydrodynamics to compensate for entropy production in the viscous evolution. In hadronic gas where $\pi^{\mu\nu}\sim0$, the shapes are constant because of the Lorentz invariance of $M$. In the QGP where $\pi^{\mu\nu}$ is significant, the shape should be changed in the region where the viscous correction to the thermal rate is large; at large invariant masses $M\sim2-3$ GeV, cf. Fig. {\ref{pic:yield_ideal_vs_visc_M_qgp}}. 
%
%\begin{figure}[!ht]
%\vspace{0.85cm}
%\begin{center}
%\includegraphics[scale=0.35]{QGP_all_inv_mass_improved.eps}
%\end{center}
%\caption{(Color online) Dilepton yield from the hadronic gas and QGP phase at minimum bias as a function of invatiant mass. }\label{pic:yield_ideal_vs_visc_M_qgp}
%\end{figure}   
%
%%%%%%%%%%%%%%%%%%%%%%%%%%%%%%%%%%%%%%%%%%%%%%%%%%%%%%%%%%%%%%%%%%%%%%%%%%%%
\subsection{Thermal dilepton elliptic flow}
%%%%%%%%%%%%%%%%%%%%%%%%%%%%%%%%%%%%%%%%%%%%%%%%%%%%%%%%%%%%%%%%%%%%%%%%%%%%
Regarding the shear viscosity and its experimental signature in relativistic heavy ion collisions, flow coefficients, for example that of elliptic flow, are more sensitive to the presence of viscosity than any particle spectra. Penetrating probes such as photons and dileptons are ideal to study viscosity, as they are influenced by the entire evolution of the medium \cite{chatterjee-srivastava-heinz-gale,Deng:2010pq,Mohanty:2011fp}. Hadrons, on the other hand, will reflect properties that  prevailed at the point of their last scattering. 

The elliptic flow of thermal lepton pairs is quantified through $v_2$, a Fourier coefficient of the azimuthal angle expansion of the yield spectrum with respect to the reaction plane 
\begin{eqnarray}
\frac{d N}{dM p_T dp_T d\phi dy} = \frac{1}{2\pi} \frac{dN}{dM p_T dp_T dy}\left\{1+\sum^\infty_{n=1}2 v_n \cos\left[n\left(\phi-\psi_r\right)\right] \right\}
\end{eqnarray}
With the averaged initial conditions used in this study, $\psi_r$ is set to zero.

%%%%%%%%%%%%%%%%%%%%%%%%%%%%%%%%%%%%%%%%%%%%%%%%%%%%%%%%%%%%%%%%%%%%%%%%%%%%
%\subsubsection{Interpretation of $v_2(p_T)$ spectra of dileptons}
%%%%%%%%%%%%%%%%%%%%%%%%%%%%%%%%%%%%%%%%%%%%%%%%%%%%%%%%%%%%%%%%%%%%%%%%%%%%

\begin{figure}[!ht]
\vspace{0.60cm}
\begin{center}
%\begin{tabular}{c c}
\includegraphics[scale=0.3]{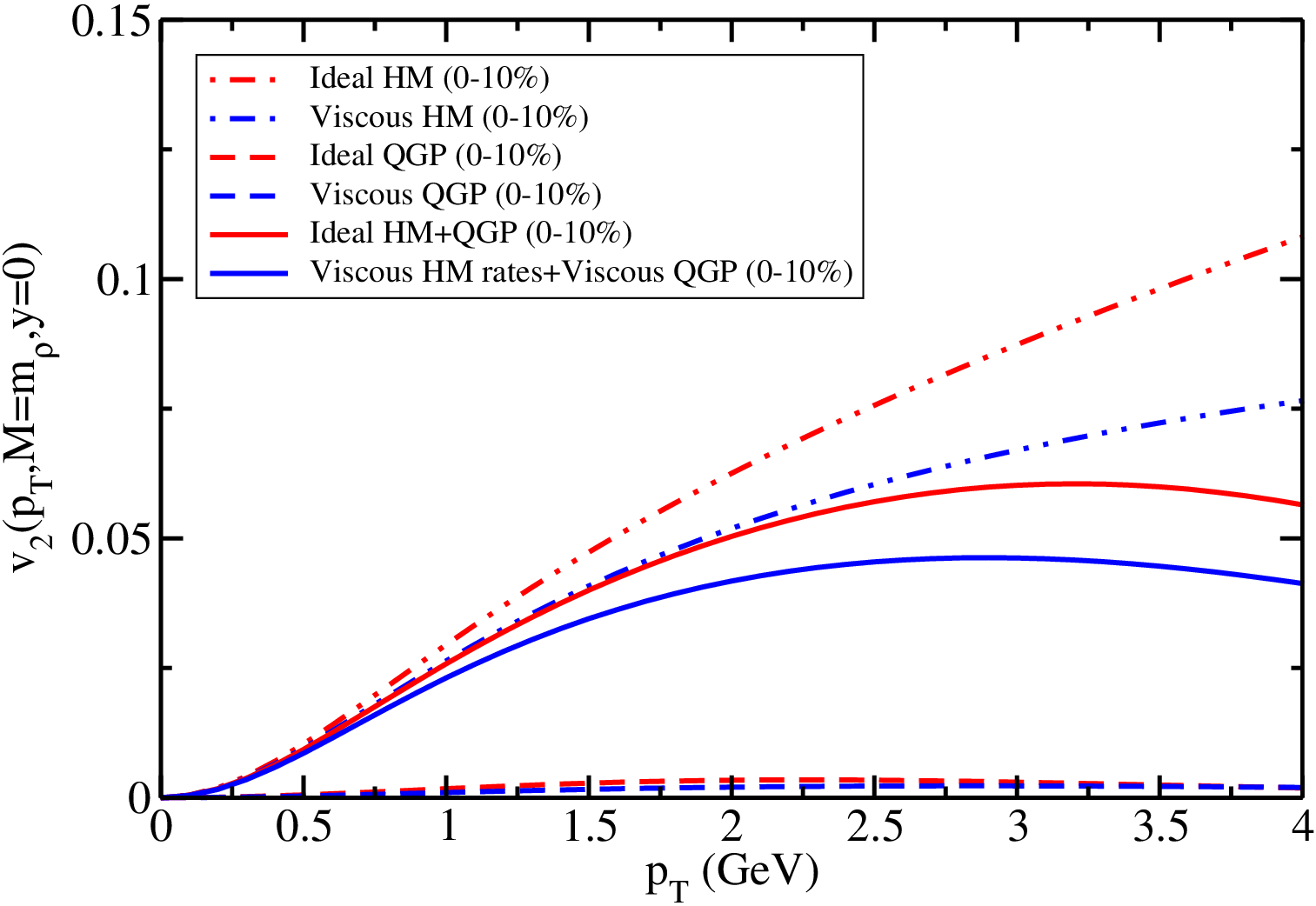} \hspace*{1cm}  \hspace{0.5cm} \includegraphics[scale=0.3]{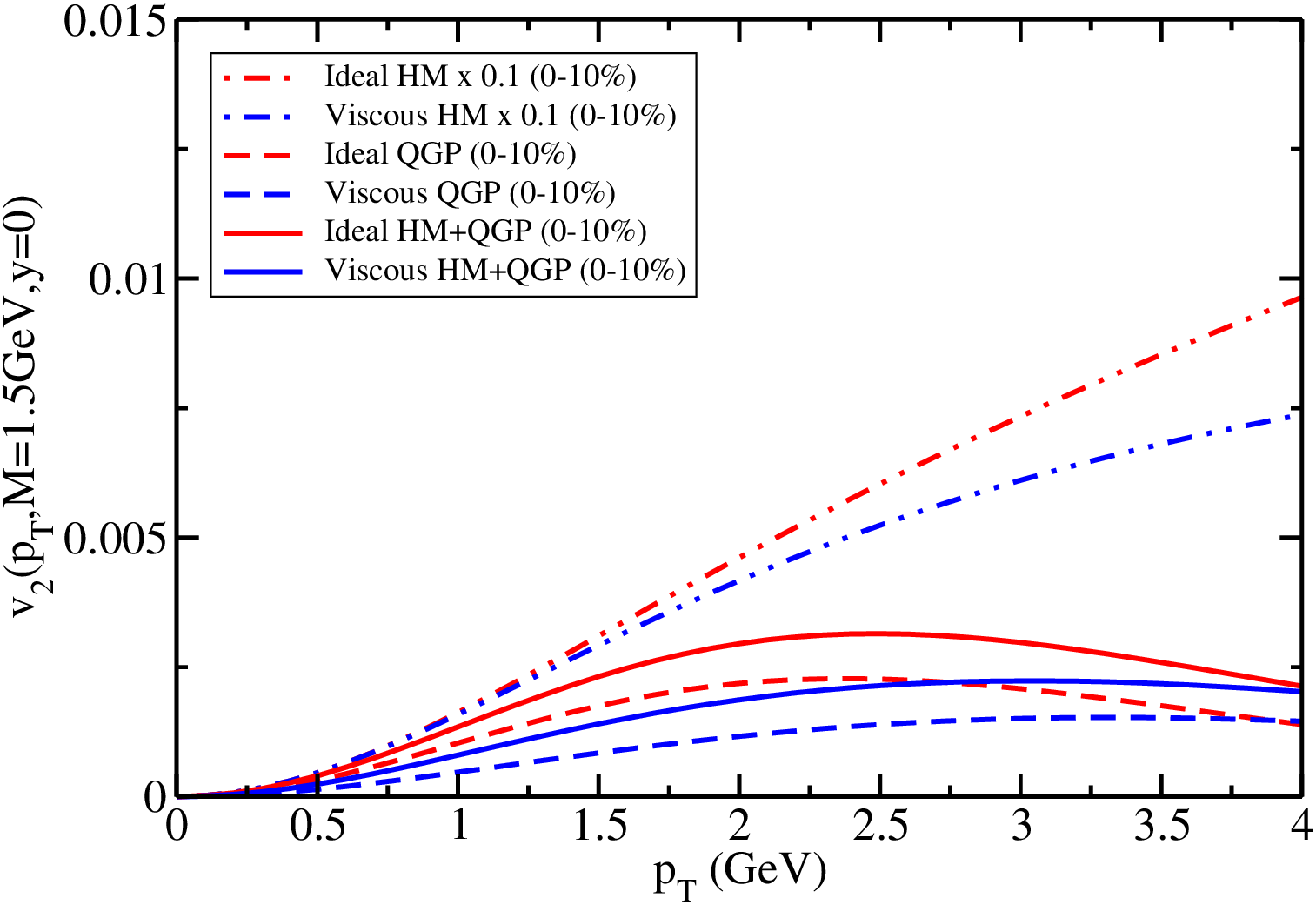}\\
%\end{tabular}
\end{center}
\caption{(Color online) Dilepton $v_2$ from the hadronic medium  and QGP  as a function of $p_T$ for two invariant masses. The panel on the left is for $M=m_\rho$, whereas the one on the right is for $M=1.5$ GeV (note the scaling applied to the HM $v_2$). The calculations shown here are for the 0 -10\% centrality class. }\label{pic:v2_ideal_vs_visc_pt_hg_qgp}
\end{figure}

Shear viscosity introduces friction between adjacent fluid layers, thus coupling faster moving fluid layers to slower moving ones, which ultimately isotropizes the angular velocity distribution of the medium and slows down its expansion.  As is the case for hadrons, the elliptic flow ($v_2$) of dileptons as a function of invariant mass is modified by the presence of shear viscosity. Following a sequence similar to that of the previous section, we start by presenting our $v_2$ results as a function of $p_T$ at fixed invariant masses \cite{chatterjee-srivastava-heinz-gale}  in Fig. \ref{pic:v2_ideal_vs_visc_pt_hg_qgp}. 

At all invariant masses, the effect of viscosity is to reduce $v_2$ of dileptons. This can be seen by comparing the red (ideal) and blue (viscous) curves in Fig. \ref{pic:v2_ideal_vs_visc_pt_hg_qgp}. Importantly, when several sources of dileptons contribute to the net dilepton yield, the final $v_2$ is a weighted average of the different elliptic flows, with the weight being the dilepton yield. This makes the interpretation of both panels of Fig. \ref{pic:v2_ideal_vs_visc_pt_hg_qgp} clear: in the low mass region, where the HM thermal dileptons outshine those from the QGP, one observes the net $v_2$ to follow more closely that of the HM. At higher invariant masses ($M=1.5$ GeV)  where the QGP yield dominates that of   the HM, the final thermal dilepton $v_2$ is close to that of the dileptons from the QGP, even though $v_2^{\rm HM} > v_2^{\rm QGP}$. Therefore, monitoring the thermal dileptons as a function of their invariant mass should help to map out the transition from a HM-dominated regime to that of a QGP. Together with a model for the time-evolution of the colliding system, such measurements could turn into a measurement of the effective temperature of the different phases. In addition, as is more clearly observed for the QGP dilepton distribution, the viscous corrections reduce the peak of $v_2$ by 45\% and shifts it to higher momenta, mainly because of the momentum-dependence of $\delta n$ . The results shown here consistently include the effects of viscosity, of using a medium-dependent vector spectral density, and of using a 3+1D hydrodynamics simulation.

\begin{center}
\begin{figure}[!htb]
\vspace{0.40cm}
\includegraphics[scale=0.35]{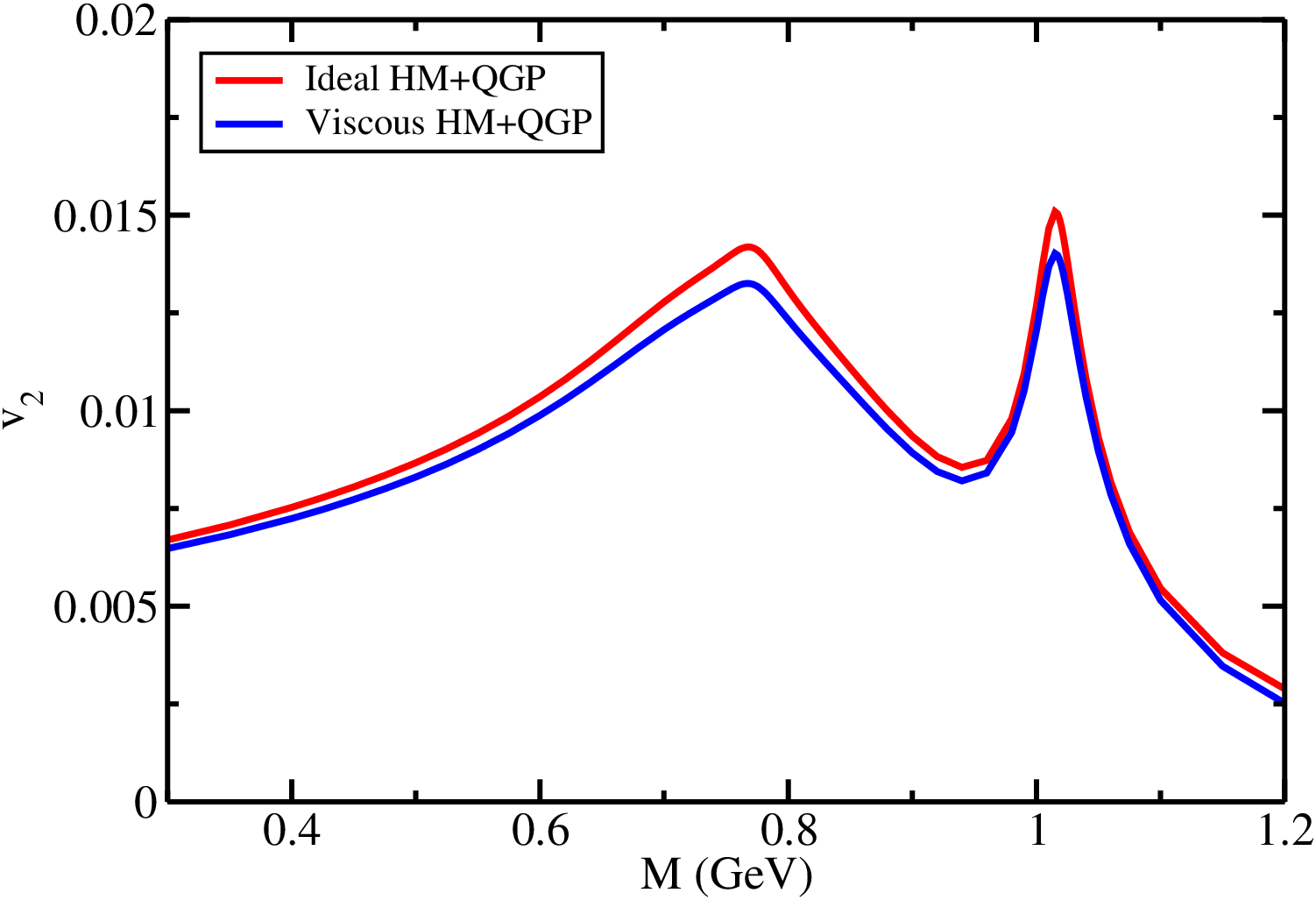} %\hspace*{1cm}  \includegraphics[scale=0.3]{v2_M_broadening_w_QGP_improved_18_sept_2013.eps}\\
\caption{(Color online) The thermal dilepton $v_2$ as a function of $M$ for both ideal hydrodynamics (top curve) and viscous hydrodynamics (bottom curve). }
\label{pic:v2_ideal_vs_visc_M_hg_qgp}
\end{figure} 
\end{center}
%%%%%%%%%%%%%%%%%%%%%%%%%%%%%%%%%%%%%%%%%%%%%%%%%%%%%%%%%%%%%%%%%%%%%%%%%%%%
%\subsubsection{Invariant mass distribution of elliptic flow}
%%%%%%%%%%%%%%%%%%%%%%%%%%%%%%%%%%%%%%%%%%%%%%%%%%%%%%%%%%%%%%%%%%%%%%%%%%%%

The distribution of $v_2$ as a function of invariant mass is given in Fig. \ref{pic:v2_ideal_vs_visc_M_hg_qgp}. There, one can clearly see that the peaks related to the $\rho-\omega$ complex and to the  $\phi$ are present in the $v_2$ spectrum | also noticed in Ref. \cite{chatterjee-srivastava-heinz-gale}. 
Unlike the invariant mass distribution of the yield, the $v_2$ distribution is actually sensitive to the presence of viscosity: it is decreased compared to its value in the inviscid case (see Fig. \ref{pic:v2_ideal_vs_visc_M_hg_qgp}). One also notices that $\rho-\omega$ complex is made slightly broader by the viscous dynamics, owing to the different temperature and flow profiles. 

The study of thermal dileptons is challenging experimentally, as competing sources have to be removed. In the intermediate mass region, the most important of these sources is charm/beauty hadrons \footnote{One reaction producing dileptons that was not included here is $4\pi\rightarrow e^+ e^-$. This channel  was found to be sub-dominant at SPS energies \protect\cite{Ruppert:2007cr}, but will be considered in the future.}. 
%Since Drell-Yan processes are of the perturbative nature, they can be removed from the data by picking out all intermediate mass dileptons that do not flow; the medium was not created during their emission. 
Charmed and beauty hadrons require precise c- and b-quark tagging before they can be removed. However, the physics of heavy flavor dileptons is interesting in and of itself, as it opens a ``clean'' window to study heavy quark energy loss and gain  mechanisms. Thus, Section \ref{charmresults} of this paper is precisely dedicated to heavy quarks, more specifically to charmed quarks.

\subsection{Including the dilepton contribution from charm decays}
\label{charmresults}
%%%%%%%%%%%%%%%%%%%%%%%%%%%%%%%%%%%%%%%%%%%%%%%%%%%%%%%%%%%%%%%%%%%%%%%%%%%%

\begin{figure}[!hb]
\vspace{0.5cm}
   \begin{center}
%      \begin{tabular}{c c}
	\includegraphics[scale=0.3]{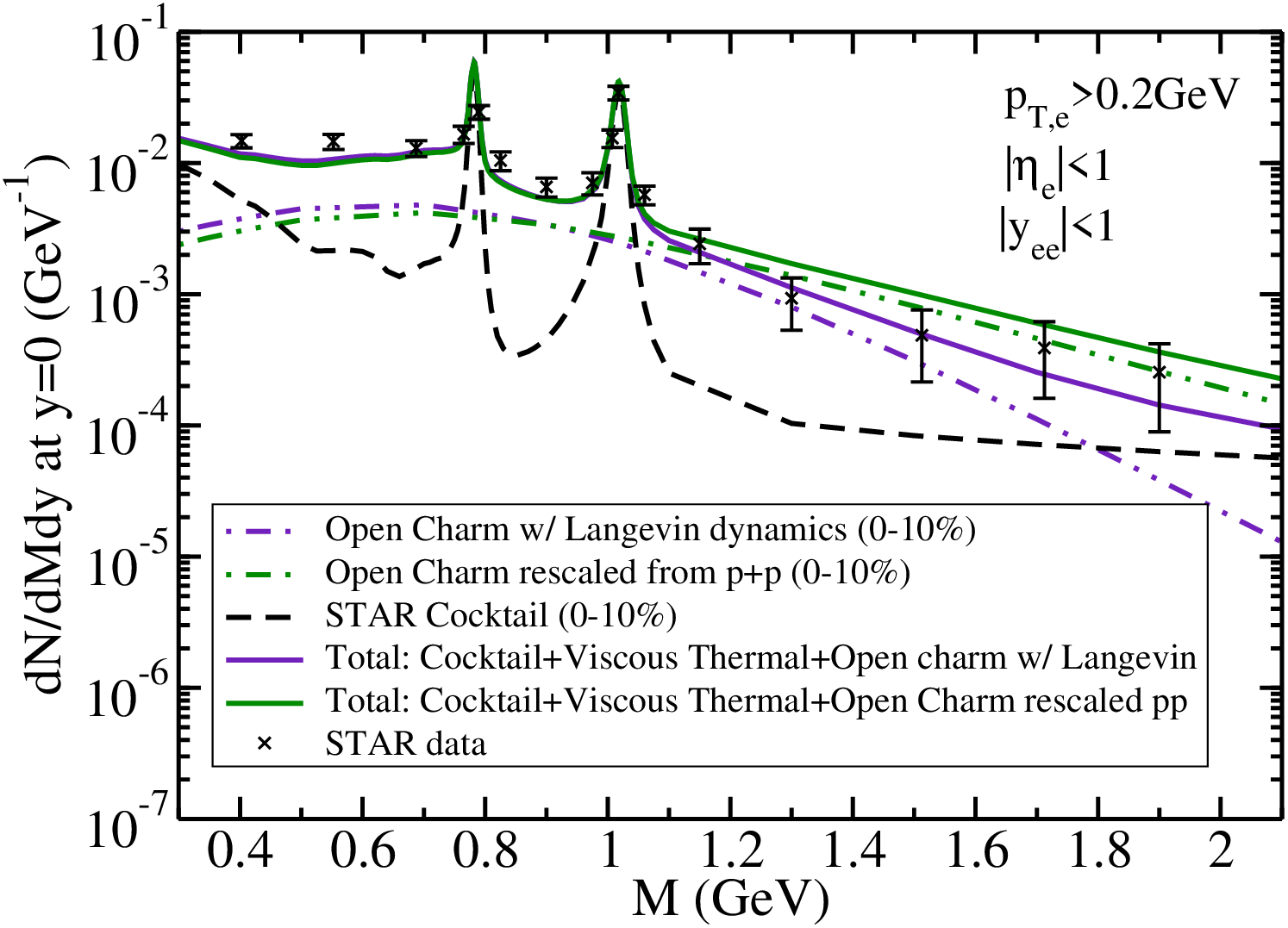} \hspace*{1cm} \includegraphics[scale=0.3]{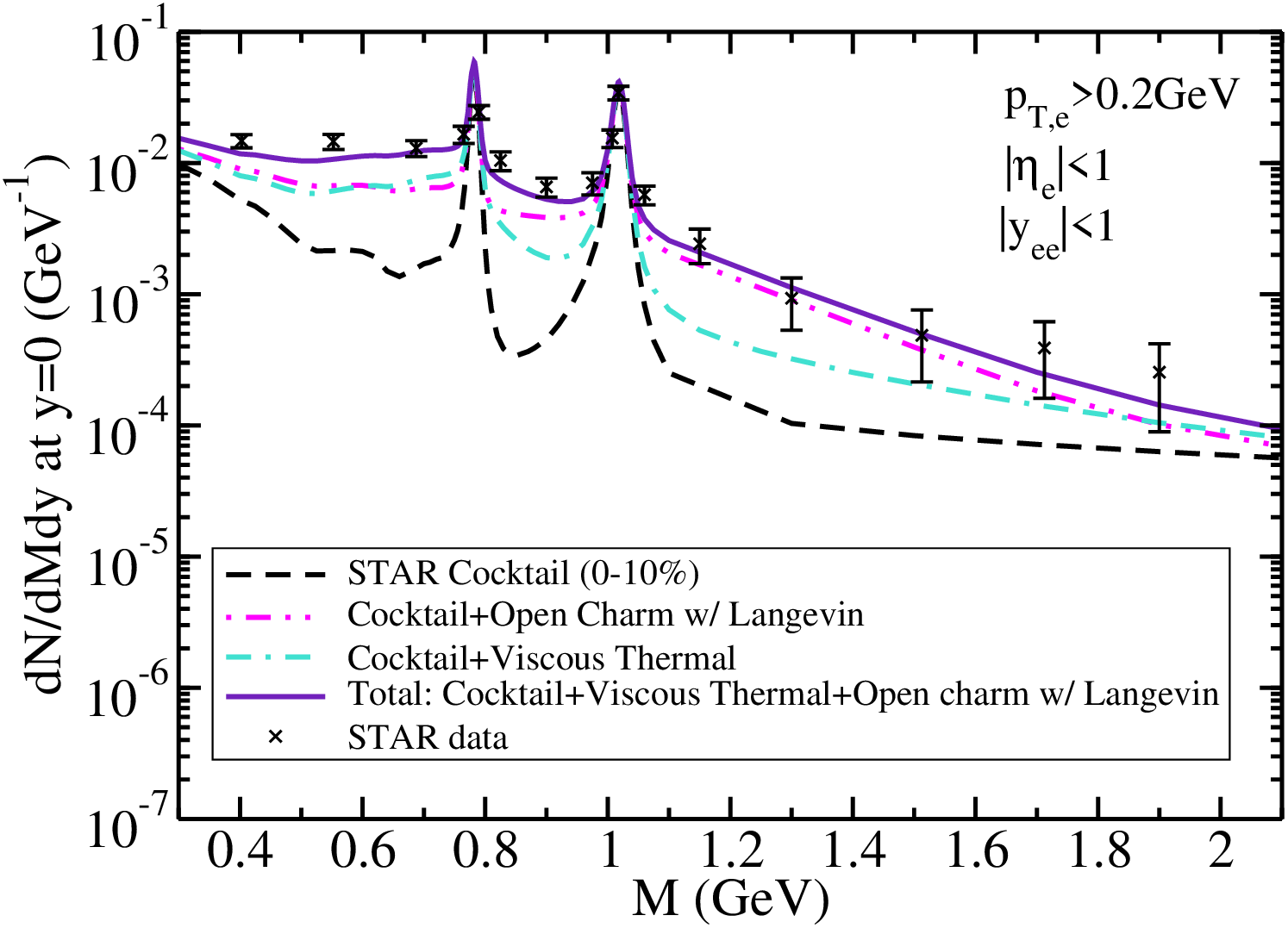}
%      \end{tabular}
   \end{center}
   \caption{Left panel: Dilepton invariant mass yields compared with experimental data at 0-10\% centrality: importance of Langevin dynamics. Right panel: Dilepton invariant mass yields compared with experimental data at 0-10\% centrality: importance of thermal radiation. The experimental acceptance cuts are indicated on the figures.}
   \label{pic:dN_dM_thermal_charm_expt_cocktail}
\end{figure}
In order to make comparisons with experimental results for dilepton yields for invariant masses up to | and including | the ``intermediate mass region'', the contribution from semi-leptonic decays of charm to dileptons must be included. As discussed in Section \ref{charm}, the dynamics of heavy quarks whose velocity $\gamma v \lesssim 1$ is approximated accurately with a relativistic Langevin equation for its momentum. We use \textsc{martini} \cite{Schenke:2009gb,Young:2011ug} as an event generator for charm quarks in heavy-ion collisions: the momenta of pairs of charm quarks are sampled using \textsc{pythia8} and the geometry in the transverse plane is sampled with the Glauber model, the Langevin equation is solved using the same calculations with \textsc{music} (including shear viscosity) that determined the thermal dilepton rates, and finally the species of charmed hadrons, and their decays, are sampled. 

The total contribution to $dN/dM$ is shown in Figure \ref{pic:dN_dM_thermal_charm_expt_cocktail} (left panel), representing the comparison of all our results with preliminary data from the STAR collaboration \cite{Ruan:2012za} for the dilepton yields in gold-gold collisions at RHIC in the 0-10\% centrality class. Note that the STAR acceptance requires the electron candidates candidates to have $|\eta^{\rm e}| < 1$ and $p_T^{\rm e} > 0.2$ GeV, and dileptons to have $| y^{\rm ee}|  < 1$.   Many $\omega$, $\rho$ and $\phi$ mesons are produced in these collisions and decay into dileptons; the data from STAR includes thermal dileptons as well as dileptons from in the decays of the many hadrons produced in heavy-ion collisions. For this reason, we include the ``cocktail" yield, as evaluated by the experimental collaboration: an extrapolation of hadron yields decaying to dilepton yields. The solid green line represents the sum of the thermal rates, the cocktail, and the contribution of charm without evolution in the medium, while the solid purple line represents the sum of the thermal rates and the cocktail with the charm contribution after evolving according to relativistic Langevin dynamics. The energy exchange of charm quarks with the medium leads to a depletion in $dN/dM$ at large $M$, and the charm contribution alone can differ by an order of magnitude at $M=2.1\;{\rm GeV}$, depending on whether Langevin evolution is considered or not. The data has a slight preference for Langevin evolution,but the size of the error precludes a stronger conclusion at this point. However, the inclusion | or not | of the possibility of charm energy variation will affect any determination of the cross sections using data for dilepton yields. At lower invariant masses, the STAR data seems compatible with this theoretical calculation. However, it is clear that acceptance-corrected data will make a much more compelling case for model compatibility.

The right panel of Figure \ref{pic:dN_dM_thermal_charm_expt_cocktail} investigates the importance of thermal radiation to describe the STAR data. In the low invariant mass region, the cocktail systematically underestimates the data and including charmed hadrons (with Langevin dynamics) is not enough to raise the calculation to the level of the measurements: the inclusion of thermal radiation is crucial. For intermediate dilepton invariant masses, the situation is less clear given STAR's current experimental uncertainties. However, the trend does suggest that thermal radiation from the QGP is present and must participate in the interpretation of the data.  

\begin{figure}
%\vspace{1.25cm}
   \begin{center}
	\includegraphics[scale=0.35]{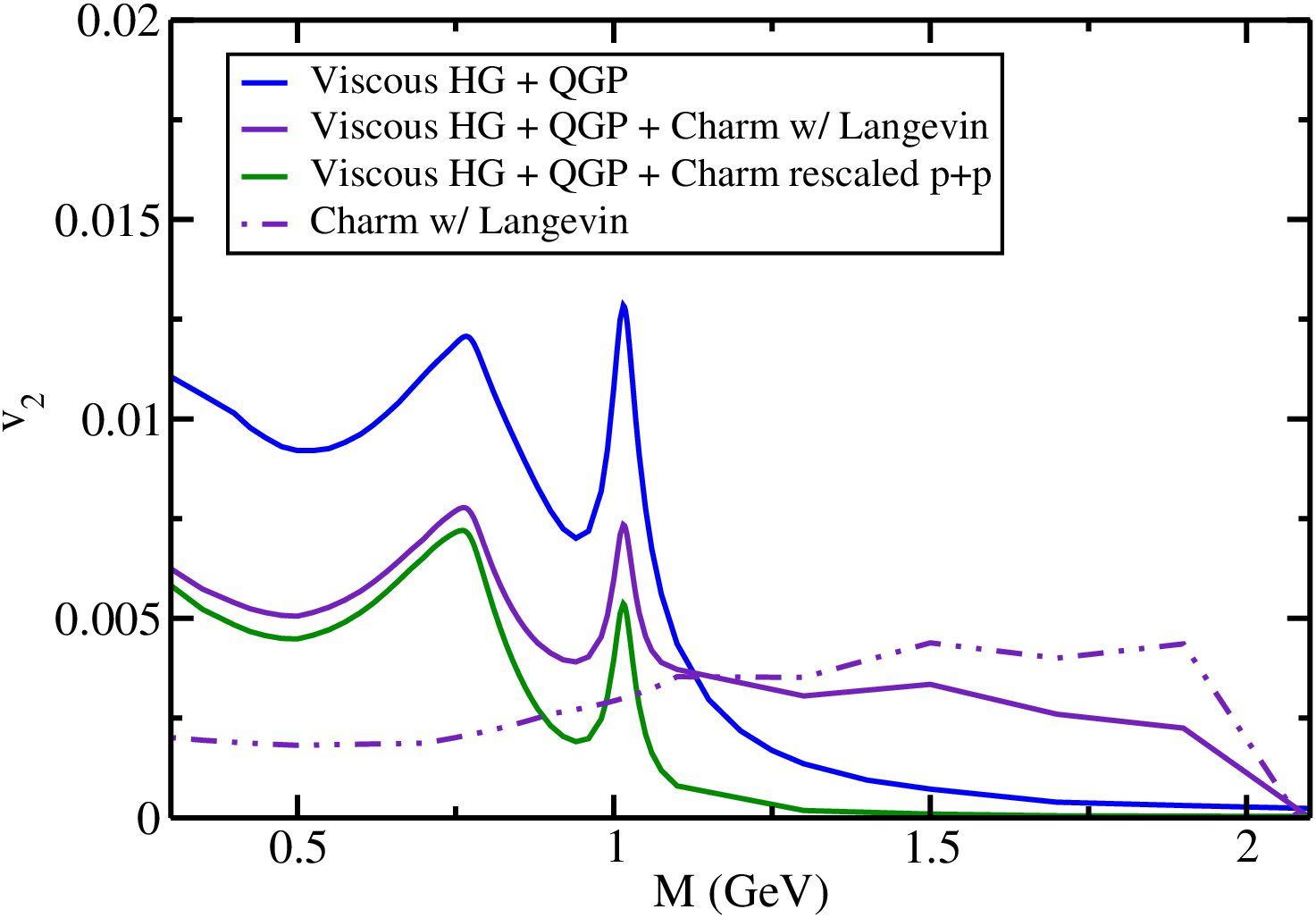}
   \end{center}
   \caption{Dilepton invariant mass $v_2$ including thermal and charm contributions at 0-10\% centrality.}
   \label{pic:v2_M_charm_b3p15}
\end{figure}

The STAR collaboration also has preliminary measurements of minimum bias $v_2(M)$ of dileptons (albeit with still large error bars) over a large momentum range, and this also includes the dileptons produced by semi-leptonic decays of charmed mesons. A comparison with these data requires knowledge of the elliptic flow of the hadronic cocktail, which we leave for a future work. The theoretical results for this observable are shown in Fig. \ref{pic:v2_M_charm_b3p15}, not including the contribution of the cocktail. %, which has non-trivial flow and will affect observables $v_n$ at the masses of the vector mesons, and also at invariant masses above about 2.5 GeV. 
Including the charm contribution to $v_2$ has two important effects: first, it reduces the $v_2$ in the 0 - 1 GeV invariant mass range by about a factor of two, and it increases the $v_2$ in the 1.5 - 2 GeV invariant mass range where the charm contribution dominates the dilepton yields. The flow of the charm contribution is smaller than the flow of the hadronic matter  contribution and it is larger than the flow of the QGP contribution, but also bear in mind that the net elliptic flow is a weighted average of its individual components. Notably, the absolute magnitude of the final elliptic flow is small. But let it be made clear again: no efforts have been made here to search for conditions that will maximize this signal, such as going to a higher centrality class, including fluctuating initial states, etc. This is left for a future systematic investigation of these effects. 

Before leaving this section on results and moving to a conclusion, it is pertinent to recall that electromagnetic radiation samples the entire space-time history of the colliding system, not just the freeze-out stage. The validity for all times of the viscosity correction linear in the viscous pressure tensor (see Eq. (\ref{delta_n})) to the thermal distribution functions can then be questioned. This investigation was performed in Ref. \cite{dion-paquet-schenke-young-jeon-gale}, those results still hold and will not be repeated here. Suffice it to say that improved versions of $\delta n$ will be explored in an upcoming work. 

%Data for $v_2(M)$ exists but does not have significant resolution in invariant mass for drawing strong conclusions at this point.

%%%%%%%%%%%%%%%%%%%%%%%%%%%%%%%%%%%%%%%%%%%%%%%%%%%%%%%%%%%%%%%%%%%%%%%%%%%%%%%%%%%%%%%%%%%%%%%%%%%%%%%%%%%%%%%%%%%%%%%%%%%%%%%%%%%%%%%%%%%%%%%%
\section{conclusion}%%%%%%%%%%%%%%%%%%%%%%%%%%%%%%%%%%%%%%%%%%%%%%%%%%%%%%%%%%%%%%%%%%%%%%%%%%%%%%%%%%%%%%%%%%%%%%%%%%%%%%%%%%%%%%%%%%%%%%%%%%%%
\label{conclusion}%%%%%%%%%%%%%%%%%%%%%%%%%%%%%%%%%%%%%%%%%%%%%%%%%%%%%%%%%%%%%%%%%%%%%%%%%%%%%%%%%%%%%%%%%%%%%%%%%%%%%%%%%%%%%%%%%%%%%%%%%%%%%%
%%%%%%%%%%%%%%%%%%%%%%%%%%%%%%%%%%%%%%%%%%%%%%%%%%%%%%%%%%%%%%%%%%%%%%%%%%%%%%%%%%%%%%%%%%%%%%%%%%%%%%%%%%%%%%%%%%%%%%%%%%%%%%%%%%%%%%%%%%%%%%%%

In this paper, we have conducted a systematic study of viscosity effects on dilepton spectra in heavy-ion collisions; (a) in the microscopic emission rates (b) in the macroscopic evolution and (c) in the semileptonic contribution. Viscosity affects the net thermal dilepton spectrum by first inducing a correction to the hadronic distribution functions. These corrections will mostly be seen in the part of the signal that is attributable to the QGP, as the shear pressure tensor, $\pi_{\mu \nu}$ is maximal in this phase. After describing the dilepton radiation in a hadronic ensemble gas and in a quark- gluon plasma and the viscous effects on the rates, those have been integrated with music, in order to consistently investigate how the viscous dynamics affects the dilepton yield and elliptic flow. Note that viscosity will also affect the cooling rate of the hydrodynamic medium, which in turn will influence both the QGP and HM thermal dileptons. For essentially all conditions considered here, the effects of the viscous dynamics are numerically not large, but are non-negligible. Moreover and importantly, the viscous corrections are required to ensure theoretical consistency.

For the purpose of comparing with recent experimental data, the calculations presented in this work include a Langevin evolution of charmed quark distributions in a viscous hydrodynamics background. The dilepton signal originating from the  charm decays was then added to that of thermal sources. These results compared well with preliminary data on Au - Au collisions from the STAR collaboration at RHIC, suggesting that the data is consistent with the viscous corrections on both microscopic rates and macroscopic dynamics. As argued previously by many authors, the intermediate invariant mass region opens a possibility to measure the energy shift of heavy quarks that interact with the hot and dense evolving medium, and the results shown here also support this assertion. Our calculations also suggest that it should be possible to access the QGP dilepton radiation in the intermediate mass region | from 1.2 GeV to 2.5 GeV | provided that precise experimental tagging of heavy flavor exists. In that case, it may be experimentally possible to remove the lepton pairs originating from open charm and beauty decays, thus exposing direct radiation from the QGP. A simultaneous analysis of yield and $v_2$ of the high-mass lepton pairs, coupled with a removal of non-photonic electrons, would produce a clear picture of the early stages of the nuclear collision. 
As written earlier in this paper, future work will include a study of varying the initial states existing prior to the hydrodynamical evolution, as well as an exploration of the effects of the different QCD transport coefficients. In what concerns measurements, the  program at RHIC together with dileptons measurements at the LHC will produce the necessary beacons of the QCD phase diagram.

%%%%%%%%%%%%%%%%%%%%%%%%%%%%%%%%%%%%%%%%%%%%%%%%%%%%%%%%%%%%%%%%%%%%%%%%%%%%%%%%%%%%%%%%%%%%%%%%%%%%%%%%%%%%%%%%%%%%%%%%%%%%%%%%%%%%%%%%%%%%%%%%  
\section*{Acknowledgments}%%%%%%%%%%%%%%%%%%%%%%%%%%%%%%%%%%%%%%%%%%%%%%%%%%%%%%%%%%%%%%%%%%%%%%%%%%%%%%%%%%%%%%%%%%%%%%%%%%%%%%%%%%%%%%%%%%%%%%
%%%%%%%%%%%%%%%%%%%%%%%%%%%%%%%%%%%%%%%%%%%%%%%%%%%%%%%%%%%%%%%%%%%%%%%%%%%%%%%%%%%%%%%%%%%%%%%%%%%%%%%%%%%%%%%%%%%%%%%%%%%%%%%%%%%%%%%%%%%%%%%%
We  are happy to acknowledge helpful discussions with G. Denicol, K. Dusling, I. Kozlov, M. Luzum, J.-F. Paquet, L. Ruan and R. Vogt. 
This work was supported in part by the Natural Sciences and Engineering Research Council of Canada, by US-DOE Contract No. DE-AC02-98CH10886,  and by US-NSF grant no. PHY-1306359. 
\bibliography{references}

%%%%%%%%%%%%%%%%%%%%%%%%%%%%%%%%%%%%%%%%%%%%%%%%%%%%%%%%%%%%%%%%%%%%%%%%%%%%%%%%%%%%%%%%%%%%%%%%%%%%%%%%%%%%%%%%%%%%%%%%%%%%%%%%%%%%%%%%%%%%%%%%
\appendix%%%%%%%%%%%%%%%%%%%%%%%%%%%%%%%%%%%%%%%%%%%%%%%%%%%%%%%%%%%%%%%%%%%%%%%%%%%%%%%%%%%%%%%%%%%%%%%%%%%%%%%%%%%%%%%%%%%%%%%%%%%%%%%%%%%%%%%
\section{Viscous corrections to QGP rates}%%%%%%%%%%%%%%%%%%%%%%%%%%%%%%%%%%%%%%%%%%%%%%%%%%%%%%%%%%%%%%%%%%%%%%%%%%%%%%%%%%%%%%%%%%%%%%%%
\label{QGPViscousCorrection}%%%%%%%%%%%%%%%%%%%%%%%%%%%%%%%%%%%%%%%%%%%%%%%%%%%%%%%%%%%%%%%%%%%%%%%%%%%%%%%%%%%%%%%%%%%%%%%%%%%%%%%%%%%%%%
%%%%%%%%%%%%%%%%%%%%%%%%%%%%%%%%%%%%%%%%%%%%%%%%%%%%%%%%%%%%%%%%%%%%%%%%%%%%%%%%%%%%%%%%%%%%%%%%%%%%%%%%%%%%%%%%%%%%%%%%%%%%%%%%%%%%%%%%%%

The ansatz for the form of the viscous correction that we chose to utilize for the QGP was previously explored by \cite{Dusling:2008xj}. This ansatz originates from the continuity requirement of the the stress-energy tensor (or the Cooper-Frye formula) across the freeze-out surface. At freeze-out, the stress-energy tensor from the hydrodynamical simulation must be matched to the one from kinetic theory. That is,
\begin{eqnarray}
T^{\mu\nu}_{\rm ideal}+\pi^{\mu\nu}=\int \frac{d^3p}{(2\pi)^3 p^0} p^\mu p^\nu \left[n(p\cdot u)+\delta n(p\cdot u)\right] \label{eq:hydro-kinet}
\end{eqnarray}
Requiring that the stress-energy tensor be continuous during the entire hydro simulation implies that the viscous correction to the equilibrium distribution function must be present in dileton production rates. For the extension to the thermal distribution, we use:      
\begin{eqnarray}
n_{\rm total}(p\cdot u) & = & n(p\cdot u) + \delta n(p\cdot u) \nonumber \\
                        & = & n(p\cdot u) + \frac{C}{2T^2(\epsilon+P)}n(p\cdot u)(1 \pm n(p\cdot u))p^\alpha p^\beta \pi_{\alpha \beta} \nonumber \\
                        & = & n(p\cdot u) + \frac{C}{2}n(p\cdot u)(1 \pm n(p\cdot u)) \frac{p^\alpha}{T} \frac{p^\beta}{T} \frac{\pi_{\alpha \beta}}{\epsilon +P}\label{eq:delta_f} 
\end{eqnarray}
where $p^\alpha$ is the 4-momentum of one of the incoming quarks, $\epsilon+P$ is the local energy density and pressure respectively, $T$ is the temperature, and $\pi_{\alpha \beta}$ is the shear-stress tensor of the fluid. Substituting Eq.(\ref{eq:delta_f}) into Eq.(\ref{eq:hydro-kinet}) yields
\begin{eqnarray}
\pi^{\mu\nu} = \left[\frac{C}{2}\int \frac{d^3p}{(2\pi)^3 p^0} n(p\cdot u)(1 \pm n(p\cdot u)) p^\mu p^\nu \frac{p^\alpha}{T} \frac{p^\beta}{T} \right] \frac{\pi_{\alpha \beta}}{\epsilon+P} \label{eq:shear-stress} 
\end{eqnarray} 
$C$ is a proportionality constant that relates the hydrodynamical shear-stress tensor to its kinetic theory counterpart. In the context of a single component ensemble, $C$ can be determined via \cite{teaney}: 
\begin{eqnarray}
\eta & = & \frac{C}{15 T^3}\int \frac{d^3 p}{(2\pi)^3p^0} n(p\cdot u)(1 \pm n(p\cdot u))\left[p^2-(u\cdot p)^2\right]^2 \label{eq:eta-c}
\end{eqnarray}
One can solve for $C$ in Eq.(\ref{eq:eta-c}) by expressing $T^3$ in terms of entropy density: 
\begin{eqnarray}
s & = & \frac{4}{3} \frac{\epsilon}{T}\\
\epsilon & = & \frac{T^4 g}{2\pi^2} \int^{\infty}_{y} \frac{x^3\sqrt{1-(y/x)^2}dx}{e^{x} \pm 1}
\end{eqnarray}
where $\epsilon$ is the average energy density of a Fermi or Bose gas with distribution $n$, $x=(p\cdot u)/T$, $y=\sqrt{p^2}/T$, $g$ is the spin degeneracy factor, and $p^2$ is the 4-momentum squared. Finally solving for $C$ is simplest in the rest frame of the fluid. 
\begin{eqnarray}
C & = & \frac{4\tilde{a}}{3\tilde{b}}\nonumber \\
\tilde{a} & = & \frac{1}{2\pi^2} \int^{\infty}_{y} dx \frac{x^3 \sqrt{1-(y/x)^2}}{e^x \mp 1} \nonumber\\
\tilde{b} & = & \frac{1}{30\pi^2} \int^{\infty}_{y} dx \frac{x^5\left[1-(y/x)^2\right]^{5/2}}{e^x \mp 1}\left\{1 \pm \frac{1}{e^x \mp 1}\right\}\label{eq:C_a}
\end{eqnarray}
For the specific case of the QGP, in the approximation of a single component fluid of massless quarks, $C$ can be evaluated analytically and is $C_q=\frac{7\pi^4}{675\zeta(5)}\approx 0.97$. 

The modification of the distribution functions owing to viscosity have a non-trivial effect on the viscous rates of QGP dileptons. Since we will be including viscous effects on the hadronic dilepton rates, it is instructive to carefully explore the manner in which the simpler Born QGP rates get modified. Indeed, we will use the same procedure for the HM case. 

In the massless quark limit, 
\begin{eqnarray}
\frac{d^4 R}{d^4 q} & = & \int \frac{d^3 p_1 d^3 p_2}{(2\pi)^6 p^0_1 p^0_2 } n(p_1\cdot u) n(p_2\cdot u) \frac{q^2}{2} \sigma \delta^4(q-p_1-p_2) \\ \nonumber
                  & + & \int \frac{d^3 p_1 d^3 p_2}{(2\pi)^6 p^0_1 p^0_2 } n(p_1\cdot u) n(p_2\cdot u) (1-n(p_1\cdot u)) \frac{q^2}{2} \sigma \delta^4(q-p_1-p_2) C_q \frac{p^\alpha_1}{T} \frac{p^\beta_1}{T}\frac{\pi_{\alpha\beta}}{\epsilon + P}\\ \nonumber
\frac{d^4 R}{d^4 q} & = & \frac{d^4 R_{ideal}}{d^4 q} + C_q \frac{J^{\alpha\beta}}{T^2} \frac{\pi_{\alpha\beta}}{\epsilon + P}
\end{eqnarray}
where we decomposed the rate into its ideal and viscous contribution ignoring all viscous corrections of order $(\delta n)^2$. Performing this integral is non-trivial. However, we know that the tensor $J^{\alpha\beta}$ of viscous correction to the rate must solely depend on the momentum of the virtual photon $q^\alpha$, the flow $u^\alpha$, and the metric $g^{\alpha\beta}$. Hence,  
\begin{eqnarray}
J^{\alpha\beta}=b_0 g^{\alpha\beta} + b_1 u^\alpha u^\beta + b_2 q^\alpha q^\beta + b_3(u^\alpha q^\beta + u^\beta q^\alpha) + b_4(u^\alpha q^\beta - u^\beta q^\alpha)
\end{eqnarray}
This is the most general form one can write down for $J^{\alpha\beta}$. However, since $J^{\alpha\beta}$ is contracted with $\pi^{\alpha\beta}$ | which must be a symmetric tensor (as part of $T^{\alpha\beta}$); any anti-symmetric piece of $J^{\alpha\beta}$ must not contribute to this calculation as shown below. The coefficients $b_0$ through $b_4$ are obtained as 
\begin{eqnarray}
\left[
\begin{array}{c}
g^{\alpha\beta}J_{\alpha\beta} \\
u^\alpha u^\beta J_{\alpha\beta} \\
q^\alpha q^\beta J_{\alpha\beta} \\
(u^\alpha q^\beta + u^\beta q^\alpha) J_{\alpha\beta}\\
(u^\alpha q^\beta - u^\beta q^\alpha) J_{\alpha\beta}
\end{array}
\right] & = & \left[\begin{array}{c c c c c}
4 & 1 & q^2 & 2(u\cdot q) & 0 \\
1 & 1 & (u\cdot q)^2 & 2(u\cdot q) & 0 \\
q^2 & (u\cdot q)^2 & q^4 & 2q^2(u\cdot q) & 0 \\
2(u\cdot q) & 2(u\cdot q) & 2q^2(u\cdot q) & 2(q^2+(u\cdot q)^2) & 0 \\
0 & 0 & 0 & 0 & 2q^2 
\end{array}
\right]
\left[\begin{array}{c}
      b_0\\
      b_1\\
      b_2\\
      b_3\\
      b_4 
      \end{array}
\right]    
\end{eqnarray}
whose solution is 
\begin{eqnarray}
\left[\begin{array}{c}
      b_0\\
      b_1\\
      b_2\\
      b_3\\
      b_4 
      \end{array}
\right] & = & \left[\begin{array}{c c c c c}
\frac{1}{2} & -\frac{1}{2}\frac{q^2}{q^2-(u\cdot q)^2} & -\frac{1}{2}\frac{1}{q^2-(u\cdot q)^2} & \frac{1}{2}\frac{u\cdot q}{q^2-(u\cdot q)^2} & 0\\
-\frac{1}{2}\frac{q^2}{q^2-(u\cdot q)^2} & \frac{3}{2}\left[\frac{q^2}{q^2-(u\cdot q)^2}\right]^2 & \frac{1}{2}\frac{q^2+2(u\cdot q)^2}{\left[q^2-(u\cdot q)^2\right]^2}& -\frac{3}{2}\frac{q^2(u\cdot q)}{\left[q^2-(u\cdot q)^2\right]^2} & 0\\
-\frac{1}{2}\frac{1}{q^2-(u\cdot q)^2} & \frac{1}{2}\frac{q^2+2(u\cdot q)^2}{\left[q^2-(u\cdot q)^2\right]^2} & \frac{3}{2}\frac{1}{\left[q^2-(u\cdot q)^2\right]^2} & -\frac{3}{2}\frac{u\cdot q}{\left[q^2-(u\cdot q)^2\right]^2} & 0\\
\frac{1}{2}\frac{u\cdot q}{q^2-(u\cdot q)^2} & -\frac{3}{2}\frac{q^2(u\cdot q)}{\left[q^2-(u\cdot q)^2\right]^2} & -\frac{3}{2}\frac{(u\cdot q)}{\left[q^2-(u\cdot q)^2\right]^2} & \frac{1}{2}\frac{q^2+2(u\cdot q)^2}{\left[q^2-(u\cdot q)^2\right]^2} & 0 \\
0 & 0 & 0 & 0 & \frac{1}{2q^2}
\end{array}
\right]
\left[
\begin{array}{c}
g^{\alpha\beta}J_{\alpha\beta} \\
u^\alpha u^\beta J_{\alpha\beta} \\
q^\alpha q^\beta J_{\alpha\beta} \\
(u^\alpha q^\beta + u^\beta q^\alpha) J_{\alpha\beta}\\
(u^\alpha q^\beta - u^\beta q^\alpha) J_{\alpha\beta}
\end{array}
\right] \nonumber
\end{eqnarray}
A simplification of the second rank tensor $J^{\alpha\beta}$ is made possible by using the identities $u^\alpha \pi_{\alpha\beta}=g^{\alpha\beta}\pi_{\alpha\beta}=0$. Indeed, $J^{\alpha\beta}$ is only proportional to $q^\alpha q^\beta$ and the proportionality constant $b_2$ is obtained via the projection operator 
\begin{eqnarray}
P_{\alpha\beta} & = & \frac{1}{2}\frac{g_{\alpha\beta}}{(u\cdot q)^2-q^2} + \frac{1}{2}\left[\frac{q^2+2(u\cdot q)^2}{\left[q^2-(u\cdot q)^2\right]^2}\right] u_{\alpha} u_{\beta} + \frac{3}{2}\frac{q_\alpha q_\beta}{\left[q^2-(u\cdot q)^2\right]^2} \nonumber\\
                & - & \frac{3}{2}\left[\frac{u\cdot q}{\left[q^2-(u\cdot q)^2\right]^2}\right]\left(u_\alpha q_\beta + u_\beta q_\alpha \right) \label{eq:projection-op}
\end{eqnarray} 
Since $P_{\alpha\beta}J^{\alpha\beta}$ is a Lorentz invariant quantity, the most efficient way to compute it is in the rest frame of the fluid cell. Performing that computation yields:
\begin{eqnarray}
b_2  =  P_{\alpha\beta}J^{\alpha\beta} & = &  \frac{1}{2|{\bf q}|^5} \int^{E_+}_{E_-} \frac{dE_1}{(2\pi)^5} \frac{q^2}{2} \sigma n(E_1)n(q^0-E_1)(1-n(E_1)) D \nonumber\\
D   & = & \left[(3q_0^2-|{\bf q}|^2) E^2_1-3 q^0 E_1 q^2 + \frac{3}{4} q^4\right]
\end{eqnarray}
where $E_{\pm}=\frac{q^0\pm |{\bf q}|}{2}$. Finally, the Born Rate with viscous corrections reads:
\begin{eqnarray}
\frac{d^4R}{d^4 q} & = & \frac{q^2}{2} \frac{\sigma}{(2\pi)^5}\left[ \frac{1}{\exp(\beta q^0)-1}\left\{1-\frac{2}{\beta |\bf q|} \ln \left[ \frac{n_{-}}{n_{+}}\right] \right\} \right. \nonumber \\
                  & + &                                      \left. C_q \frac{q^\alpha}{T} \frac{q^\beta}{T} \frac{\pi_{\alpha\beta}}{\epsilon + P} \frac{1}{2|{\bf q}|^5} \int^{E_+}_{E_-} dE_1 n(E_1)n(q^0-E_1)(1-n(E_1)) D \right]
\end{eqnarray}
% 
%%%%%%%%%%%%%%%%%%%%%%%%%%%%%%%%%%%%%%%%%%%%%%%%%%%%%%%%%%%%%%%%%%%%%%%%%%%%%%%%%%%%%%%%%%%%%%%%%%%%%%%%%%%%%%%%%%%%%%%%%%%%%%%%%%%%%%%%%%%%%%%%
\section{The vector meson self-energy and its viscous correction}%%%%%%%%%%%%%%%%%%%%%%%%%%%%%%%%%%%%%%%%%%%%%%%%%%%%%%%%%%%%%%%%%%%%%%%%%%%%%%%
\label{HGViscousCorrection}%%%%%%%%%%%%%%%%%%%%%%%%%%%%%%%%%%%%%%%%%%%%%%%%%%%%%%%%%%%%%%%%%%%%%%%%%%%%%%%%%%%%%%%%%%%%%%%%%%%%%%%%%%%%%%%%%%%%%
%%%%%%%%%%%%%%%%%%%%%%%%%%%%%%%%%%%%%%%%%%%%%%%%%%%%%%%%%%%%%%%%%%%%%%%%%%%%%%%%%%%%%%%%%%%%%%%%%%%%%%%%%%%%%%%%%%%%%%%%%%%%%%%%%%%%%%%%%%%%%%%%
Using the tools of the previous section, the goal of this section is to derive the viscous correction to the self-energy. To this end, we first outline the steps leading to the thermal self-energy, and then we extend it to include viscous corrections.  

\subsection{Thermal self-energy}

To simplify the calculation, and without loss of generality, we choose the z-axis such that the 4-momentum of particle $V$ is aligned with it, i.e. $p^\mu=(E,0,0,|{\bf p}|)$. We further define the angle $\theta$ between the z-axis and the momentum $k^\mu=(\omega,{\bf k})$ of particle $a$. Note that $\theta$ is {\it not} the angle between $p^\mu$ and $k^\mu$.  

In the rest frame of particle $a$, it is possible to evaluate the angular part of the self-energy integral. From now on, prime ($'$) is used to denote energy and momentum in $V$'s rest frame and double prime ($''$) is used to label $a$'s rest frame. One can relate the energy in the two frames via:
\begin{eqnarray}
s = m^2_V + m^2_a + 2E''m_a = m^2_V + m^2_a + 2m_V\omega'
\end{eqnarray}
Hence, $E''=\frac{m_V}{m_a}\omega'$. Furthermore, in $V$'s rest frame, $\omega=\frac{E\omega'+|{\bf p}||{\bf k'}|z'}{m_V}$, where $z'=\cos\theta'$. Putting everything together,
\begin{eqnarray}
\Pi^{\rm T}_{Va}(|{\bf p}|, T) &=& -4 \pi \int \frac{d^3k}{(2\pi)^3\omega} n_a(\omega) \sqrt{s}f^{\rm c.m.}_{Va}(s)\nonumber\\
                	      &=& -4 \pi \int \frac{|{\bf k'}|^2d|{\bf k'}| dz'}{(2\pi)^2 \omega'} n_a\left(\frac{E\omega'+|{\bf p}||{\bf k'}|z'}{m_V}\right) f^{\rm a's\text{ }rest}_{Va}\left(\frac{m_V}{m_a} \omega' \right)\nonumber\\
                 	      &=& -\frac{m_V}{\pi}\int^{\infty}_{m_a}|{\bf k'}| d\omega' f^{\rm a's\text{ }rest}_{Va}\left(\frac{m_V}{m_a} \omega' \right) \int^1_{-1} dz' n_a\left(\frac{E\omega'+|{\bf p}||{\bf k'}|z'}{m_V}\right) \nonumber\\
                              &=& -\frac{m_V m_a T}{\pi |{\bf p}|}\int^{\infty}_{m_a} d\omega'\ln\left[\frac{1\pm\exp\left(-\omega_{+}/T\right)}{1\pm\exp \left(-\omega_{-}/T\right)}\right]f^{\rm a's\text{ }rest}_{Va}\left(\frac{m_V}{m_a} \omega' \right)
\end{eqnarray}
where $\omega_\pm = \frac{E\omega' \pm |{\bf p}||{\bf k'}|}{m_V}$. This expression for the self energy is evaluated on the mass shell of the vector meson $V$. 

\subsection{Viscous corrections to the thermal self-energy}

To calculate the viscous correction to the thermal self-energy, we proceed by including the $\delta n$ correction to the thermal distribution function. Unlike the bose distribution function present in the rates | which originates from the KMS relation and therefore is not related to the thermal distribution function of vector mesons | the distribution function present in the self-energy Eq.(\ref{eq:thermal_self-energy}) is indeed a distribution function of thermal particles. So the viscous correction to the thermal distribution in Eq.(\ref{eq:delta_f}) applies. Thus,  
\begin{eqnarray}
\delta \Pi^{\rm T}_{Va}(|{\bf p}|,T) = -4 \pi \int \frac{d^3k}{(2\pi)^3\omega} \delta n_a(k\cdot u) \sqrt{s}f^{\rm c.m.}_{Va}(s) =  C_a \frac{K^{\alpha\beta}}{T^2}\frac{\pi_{\alpha\beta}}{\epsilon + P}
\end{eqnarray}
Note that $C_a$ cannot be computed via Eq.(\ref{eq:C_a}), since $\delta \Pi^T_{Va}$ is describing a multi-component mixture. Hence, a simplifying assumption is made: $\forall a C_a=1$. Now we expand the tensor $K^{\mu\nu}$ in the same manner as the QGP $J^{\mu\nu}$ tensor encountered earlier:
\begin{eqnarray}
K^{\mu\nu}=B_0 g^{\alpha\beta} + B_1 u^\alpha u^\beta + B_2 p^\alpha p^\beta + B_3(u^\alpha p^\beta + u^\beta p^\alpha) + B_4(u^\alpha p^\beta - u^\beta p^\alpha)
\end{eqnarray}
Since the relation $u^\alpha \pi_{\alpha\beta}=g^{\alpha\beta}\pi_{\alpha\beta}=0$ still holds, we use the same projection operator as in Eq.(\ref{eq:projection-op}) to determine $B_2$. Thus, 
\begin{eqnarray}
B_{2,Va} & = & P_{\alpha\beta}K^{\alpha\beta} \nonumber \\
        & = & -4 \pi \int \frac{d^3k}{(2\pi)^3} n_a(u\cdot k)(1\pm n_a(u\cdot k)) \frac{\sqrt{s}}{\omega}f_{Va}(s)\left[ \frac{1}{2}\frac{m^2_a}{(u\cdot p)^2-p^2} + \frac{1}{2}\left[\frac{p^2+2(u\cdot p)^2}{\left[p^2-(u\cdot p)^2\right]^2}\right] (u\cdot k)^2  \right. \nonumber\\
                & + & \left. \frac{3}{2}\frac{(p \cdot k)^2}{\left[p^2-(u\cdot p)^2\right]^2} - 3\frac{(u \cdot p)(u \cdot k)(p \cdot k)}{\left[p^2-(u\cdot p)^2\right]^2} \right] 
\end{eqnarray}
Throughout this appendix, the upper (lower) sign refers to Bosons (Fermions). In the rest frame of the medium (using $z=\cos\theta$ as before): 
\begin{eqnarray}
B_{2,Va} & = & -4 \pi \int \frac{d^3k}{(2\pi)^3\omega} n_a(1\pm n_a)  \sqrt{s}f_{Va}\left[ \frac{m^2_a}{2|{\bf p}|^2} + \left(\frac{3E^2}{2|{\bf p}|^4}-\frac{1}{2|{\bf p}|^2}\right)\omega^2 + \frac{3}{2}\frac{(E\omega - |{\bf p}||{\bf k}|z)^2}{|{\bf p}|^4} \right. \nonumber \\
        & - & \left. \frac{3E\omega((E\omega - |{\bf p}||{\bf k}|z)}{|{\bf p}|^4} \right]\nonumber \\
        & = & -4 \pi \int \frac{d^3k}{(2\pi)^3\omega} n_a(1\pm n_a)  \sqrt{s}f_{Va}\left[ \frac{m^2_a}{2|{\bf p}|^2} +\frac{3|{\bf k}|^2 z^2 - \omega^2}{2|{\bf p}|^2}\right]
\end{eqnarray}
Evaluating the integral in the rest frame of $a$, we obtain:
\begin{eqnarray}
B_{2,Va} & = & -4 \pi m_a \int \frac{d^3k'}{(2\pi)^3\omega'} n_a\left(\frac{E\omega'+|{\bf p}||{\bf k'}|z'}{m_V}\right)\left[1\pm n_a\left(\frac{E\omega'+|{\bf p}||{\bf k'}|z'}{m_V}\right)\right]f^{\mathrm{a's\text{ }rest}}_{Va}\left( \frac{m_V}{m_a} \omega' \right)\nonumber\\ 
   & \times & \left[ \frac{m^2_a}{2|{\bf p}|^2} + \frac{3\left(\frac{E |{\bf k'}|z'+|{\bf p}|{\bf \omega'}}{m_V}\right)^2 - \left(\frac{E\omega'+|{\bf p}||{\bf k'}|z'}{m_V}\right)}{2 |{\bf p}|^2}\right]\nonumber\\
   & = & -\frac{m_a}{2\pi|{\bf p}|^2}\int^{\infty}_{m_a} d\omega'|{\bf k'}|f^{\mathrm{a's\text{ }rest}}_{Va}\left( \frac{m_V}{m_a} \omega' \right) \int^1_{-1} dz' n_b\left(\frac{E\omega'+|{\bf p}||{\bf k'}|z'}{m_V}\right)\left[1\pm n_a\left(\frac{E\omega'+|{\bf p}||{\bf k'}|z'}{m_V}\right)\right] \nonumber\\
   & \times & \left[ m^2_a + (3|{\bf p}|^2-E^2)\frac{\omega'^2}{m^2_V} + 4 E|{\bf p}|\frac{\omega'|{\bf k'}|}{m^2_V}z' + (3E^2-|{\bf p}|^2)\frac{|{\bf k'}|^2}{m^2_V} z'^2\right]
\end{eqnarray}
where $|{\bf k}|z=\frac{E}{m_V}|{\bf k'}|z'+\frac{|{\bf p}|}{m_V}\omega'$. Performing the angular integral yields:
\begin{eqnarray}
\label{B2}
B_{2,Va} & = &-\frac{m_V}{2\pi|{\bf p}|^2}\int^{\infty}_{m_a} d\omega'|{\bf k'}|f^{\mathrm{a's\text{ }rest}}_{Va}\left( \frac{m_V}{m_a} \omega' \right) \times \left( {\cal A} + {\cal B} +{\cal  C} +{\cal  D} +{\cal  E} \right)
\end{eqnarray}
where
\begin{eqnarray}                 
{\cal A} &=& \left(\frac{m_VT}{|{\bf p}||{\bf k'}|}\right)\left[m^2_a+\frac{\left(E|{\bf k'}|-|{\bf p}|\omega\right)^2-\left(E\omega'-|{\bf p}||{\bf k'}|\right)^2}{m^2_V}\right]\left[\exp(\omega_{-}/T)\mp1\right]^{-1}\nonumber\\
{\cal B} &=&-\left(\frac{m_VT}{|{\bf p}||{\bf k'}|}\right)\left[m^2_a+\frac{\left(E|{\bf k'}|+|{\bf p}|\omega\right)^2-\left(E\omega'+|{\bf p}||{\bf k'}|\right)^2}{m^2_V}\right]\left[\exp(\omega_{+}/T)\mp1\right]^{-1}\nonumber\\ 
{\cal C} &=& \pm2\left(\frac{m_VT}{|{\bf p}||{\bf k'}|}\right)^2\left[(3E^2-|{\bf p}|^2)\frac{|{\bf k'}|^2}{m^2_V} + 2\frac{E\omega'|{\bf p}||{\bf k'}|}{m^2_V}\right]\ln\left[1\mp \exp(-\omega_{+}/T)\right]\nonumber\\
{\cal D} &=& \pm2\left(\frac{m_VT}{|{\bf p}||{\bf k'}|}\right)^2\left[(3E^2-|{\bf p}|^2)\frac{|{\bf k'}|^2}{m^2_V} - 2\frac{E\omega'|{\bf p}||{\bf k'}|}{m^2_V}\right]\ln\left[1\mp \exp(-\omega_{-}/T)\right]\nonumber\\
{\cal E} &=& \mp2\left(\frac{m_VT}{|{\bf p}||{\bf k'}|}\right)^3\left[(3E^2-|{\bf p}|^2)\frac{|{\bf k'}|^2}{m^2_V}\right]\left\{ \text{Li}_2\left[\pm\exp(-\omega_+/T)\right] - \text{Li}_2\left[\pm\exp(-\omega_-/T)\right] \right\}   \ ,         
\end{eqnarray}
and Li$_2$ is the dilogarithm function. Thus, the total self-energy is
\begin{eqnarray}
\Pi^{\rm tot}_{V} (M, |{\bf p}|, T) & = & \Pi^{\rm vac}_V\left( M \right) \\
                                 & + & \sum_{a=N,\bar{N},\pi}\left\{-\frac{m_V m_a T}{\pi |{\bf p}|}\int^{\infty}_{m_a} d\omega'\ln\left[\frac{1\pm\exp\left(-\omega_{+}/T\right)}{1\pm\exp \left(-\omega_{-}/T\right)}\right]f^{\rm a's\text{ }rest}_{Va}\left(\frac{m_V}{m_a} \omega' \right) \right. \nonumber \\
                                 & + & \left. C_a B_{2,Va} \frac{p^\alpha_V p^\beta_V}{T^2} \frac{\pi_{\alpha\beta}}{\epsilon + P}\right\}\nonumber
\end{eqnarray}

\end{document}